%% file: 00_main.tex
\documentclass{vgtc}                          % final (conference style)

\ifpdf%                                % if we use pdflatex
  \pdfoutput=1\relax                   % create PDFs from pdfLaTeX
  \usepackage{graphicx}                % allow us to embed graphics files
  \DeclareGraphicsExtensions{.pdf,.png,.jpg,.jpeg} % for pdflatex we expect .pdf, .png, or .jpg files
\else%                                 % else we use pure latex
  \usepackage{graphicx}                % allow us to embed graphics files
  \DeclareGraphicsExtensions{.eps}     % for pure latex we expect eps files
\fi%

\graphicspath{{figures/}{pictures/}{images/}{./}} % where to search for the images

\usepackage{microtype}                 % use micro-typography (slightly more compact, better to read)
\PassOptionsToPackage{warn}{textcomp}  % to address font issues with \textrightarrow
\usepackage{textcomp}                  % use better special symbols
\usepackage{mathptmx}                  % use matching math font
\usepackage{times}                     % we use Times as the main font
\usepackage{cite}                      % needed to automatically sort the references
\usepackage{tabu}                      % only used for the table example
\usepackage{booktabs}                  % only used 
\usepackage{pdflscape}
\usepackage{afterpage}
\usepackage{array}
\usepackage{ragged2e}

\usepackage{subfigure}

\input{macros}

\onlineid{123}

\vgtccategory{Research}

\vgtcinsertpkg

\title{An EEG-based Experiment on VR Sickness and Postural Instability While Walking in Virtual Environments}

\author{Carlos Alfredo Tirado Cortes \\ %
        \parbox{1.3in}{\scriptsize \centering iCinema Research Centre\\ University of New South Wales\\ Randwick, Australia} %
\and Chin-Teng Lin \\ %
     \parbox{1.3in} {\scriptsize \centering Australian AI Institute\\ University of Technology, Sydney\\ Ultimo, Australia} %
\and Tien-Thong Nguyen Do \\ %
     \parbox{1.3in} {\scriptsize \centering Australian AI Institute\\ University of Technology, Sydney\\ Ultimo, Australia} %
\and Hsiang-Ting Chen \thanks{e-mail: tim.chen@adelaide.edu.au}\\ %
     \parbox{1.3in}{\scriptsize \centering University of Adelaide  \\Adelaide, Australia}}

 \CCScatlist{
   \CCScatTwelve{Human-centered computing}{Human computer interaction (HCI)}{HCI design and evaluation methods}{User studies};
   \CCScatTwelve{Human-centered computing}{Human computer interaction (HCI)}{Interaction paradigms}{Virtual reality}
 }

\abstract{
Previous studies showed that natural walking reduces the susceptibility to VR sickness.
However, many users still experience VR sickness when wearing VR headsets that allow free walking in room-scale spaces. 
This paper studies VR sickness and postural instability while the user walks in an immersive virtual environment using an electroencephalogram (EEG) headset and a full-body motion capture system. 
The experiment induced VR sickness by gradually increasing the translation gain beyond the user's detection threshold. 
A between-group comparison between participants with and without VR sickness symptoms found \toremove{no} \newtext{some} significant difference\newtext{s} in postural stability \newtext{but found none on} \toremove{or} gait patterns during the walking. 
\toremove{However}\newtext{In the EEG analysis}, the group with VR sickness showed a reduction of alpha power, a phenomenon previously linked to a higher workload and efforts to maintain postural control. 
In contrast, the group without VR sickness exhibited brain activities linked to fine cognitive-motor control. 
The EEG result provides new insights into the postural instability theory: participants with VR sickness could maintain their postural stability at the cost of a higher cognitive workload.
Our result also indicates that the analysis of lower-frequency power could complement behavioural data for continuous VR sickness detection in both stationary and mobile VR setups. 
}

\begin{document}
\maketitle

\input{10_intro}

\input{20_related_work}

\input{30_experiment}
\input{40_measurements}
\input{50_result}
\input{60_discussion_v2}

\input{70_conclusion}
% == REFERENCE
\input{71_EEG_Table}
\bibliographystyle{abbrv-doi}
\bibliography{vr_ref}
\appendix

\section{Appendix}
\input{72_EEG_Pipeline}
\input{73_EEG_ERSP}
%\input{74_EEG_Freqs}

% == BIOGRAPHY SECTION

% \begin{IEEEbiography}[{\includegraphics[width=1in,height=1.25in,clip,keepaspectratio]{Bio_Tim.jpg}}]{Hsiang-Ting Chen}
% received his Ph.D. degree in Computer Science from the National Tsing Hua University, Taiwan in 2012. He then worked as a researcher in the Igarashi Lab, University of Tokyo from 2012 to 2013 and a post-doc research fellow at Hasso Plattner Institute, Germany from 2014 to 2015. In 2016, he took a lecturer position at University of Technology Sydney and became a core member of the Center of Artificial Intelligence. He works in the field of Human Computer Interaction and Computer Graphics. He is particularly interested in creating novel interaction systems where humans team up with computers to collaboratively produce creative contents.
% \end{IEEEbiography}


%\newpage
\vfill

\end{document}

%% file: macros.tex
\usepackage{xcolor}
\usepackage{todonotes}
\usepackage{graphicx}
\usepackage{csquotes}
\usepackage{booktabs}
\usepackage{soul}

\newcommand{\toremove}[1]{}
\newcommand{\newtext}[1]{{#1}}

\newcommand{\novrs}[0]{\textit{NoVRS} }
\newcommand{\vrs}[0]{\textit{VRS} }

%% file: 10_intro.tex
\begin{figure}
    \centering
    \includegraphics[width=.75\columnwidth]{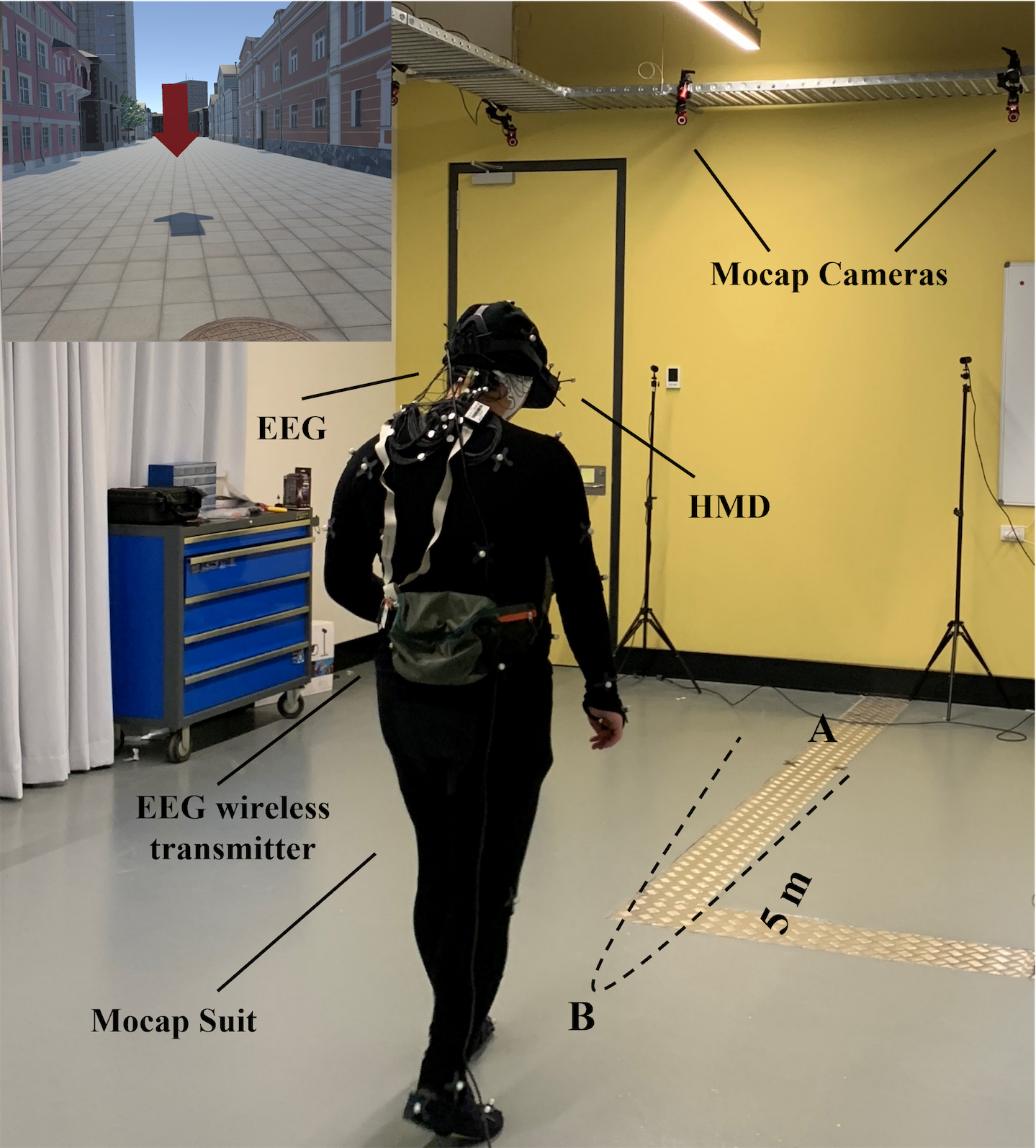}
    \caption{The experiment space and 
    the motion capture system. The participant wore a motion capture suit, a 64-channel EEG headset, and a VR headset. The participant performs a VR walking task while physically walking between points A and B in each trial. Top-left corner shows the participant's view in VR.}
    \label{fig:teaser}
\end{figure}

\section{Introduction}\label{sec:introduction}

Virtual Reality (VR) has become a mainstream product in the past few years. A recent report \cite{Alsop2021} found that six million VR headsets were sold worldwide in 2021 and evaluated the VR market size at 4.8 billion dollars.
Despite the rapid growth of the VR market, VR sickness has long been considered a major issue that prevents the wider acceptance of VR technology. 
VR sickness shares similar symptoms to those of motion sickness. However, VR sickness presents stronger disorientation symptoms than other common symptoms such as headache, nausea, vomiting, \newtext{and} drowsiness \toremove{, and disorientation} \cite{Stanney1997, LaViola2000}. 
\newtext{In 2016, a} \toremove{An early} review article \cite{Rebenitsch2016} reported that between 30\% to 80\% of head-mounted display (HMD) users experienced some VR sickness symptoms. 

The introduction of room-scale tracking to consumer VR headsets is a significant milestone in the fight against VR sickness. 
According to the sensory conflict theory, the mismatch between sensory inputs, e.g., visual and vestibular systems during passive movement, is the main source of VR sickness \cite{Rebenitsch2016, LaViola2000}.
Room-scale tracking allows users to navigate and interact with the virtual environment with natural walking, which reduces the sensory conflicts and the VR sickness symptoms \cite{Nilsson2014}.  % Jaeger2001, Bozgeyikli2016
However, even in a well-controlled lab environment, research studies still reported the occurrence of VR sickness symptoms despite using headsets and applications that allow natural walking \cite{Kruse2018, Feigl2019}.
Indeed, recent studies found that VR sickness is still a prevalent human factor issue in modern consumer VR headsets~\cite{Yildirim2020-commercial} , and around 5\% of users had experienced severe VR symptoms~\cite{Stanney2020a}.
Understanding the effect of VR sickness and its underlying cognitive and neural mechanisms is imperative to bring a comfortable VR experience for more users.

% There is limited research investigating how and when VR sickness occurs in a mobile VR setup, i.e. the user navigates the virtual environment with natural walking in the physical world. 
The VR community has accumulated a large volume of research on VR sickness \cite{Weech2019, Saredakis2020, Jasper2020}.
The vast majority of these work used a stationary experiment setup \cite{Saredakis2020} to study VR sickness.
A common protocol is to keep the participant in the sitting posture and watch 360$^{\circ}$ videos or 2D abstract scenes with strong optical flows (Table 4 in \cite{Saredakis2020}).
Some works conducted EEG-based studies to understand the brain dynamics related to VR sickness (Table \ref{tab:EEG}). 
These works further required participants to minimize head movements and stay in a stationary sitting posture to avoid introducing noise into the EEG signals.
As most modern VR headsets allow users to navigate the virtual environment with natural walking, there is a knowledge gap about VR sickness in a mobile VR setup.
% This paper focuses on using electroencephalogram(EEG)-based methods to understand VR sickness. 
% Few EEG-based studies have systematically investigated how VR sickness affects users in a mobile VR setup.

This paper aims to understand how VR sickness affects the user in a mobile VR setup using EEG and full-body motion capture. 
Our experiment follow the protocol in a previous work \cite{Cortes2019_TG}, which induces VR sickness by gradually increasing the level of translation gain (TG), i.e., the scale of mapping between the user's motions in the virtual world and those in the physical world \cite{Wilson2018, steinicke2009_threshold}.
As the TG level increases beyond the detection threshold \cite{Steinicke2010, Suma2012}, the perceived disparity between the virtual and physical worlds would also increase. 
The paper~\cite{Cortes2019_TG} showed that the increased disparity would eventually cause postural instability and VR sickness. 
% cause the loss of postural stability and force users' constant adjustment of body control which, according to the postural instability theory, would lead to VR sickness. 

In this paper we hypothesise that, rather than sensory conflicts, postural instability is a higher driver for VR sickness in mobile VR setups. 
The hypothesis is based on two previous findings: first, previous studies suggested that mobile VR decreases sensory conflicts \cite{Janeh2017, Nilsson2018}. % TODO: Removed Harris2000
Second, previous studies suggested that wearing a VR headset in a standing position would produce postural instability, \cite{Horlings2009, Chiarovano2015, Robert2016}, and the differences in biomechanics between locomotion in the virtual and the real world would also lead to the loss of body control \cite{Robert2016, Janeh2017}.
We believe that the diminished sensory conflict cues and increased efforts in postural control in mobile VR setups make the postural instability theory~\cite{Riccio1991:PIT} intriguing in the mobile VR context.
\toremove{Indeed, previous works reported evidence both for and against the postural instability theory.}
% \cite{palmisano2020_postural}. % TODO: Chiarovano2015 candidate to remove
We hope this EEG-based study will provide new insights and data to this debate.

\begin{figure}
    \centering
    \includegraphics[width=\columnwidth]{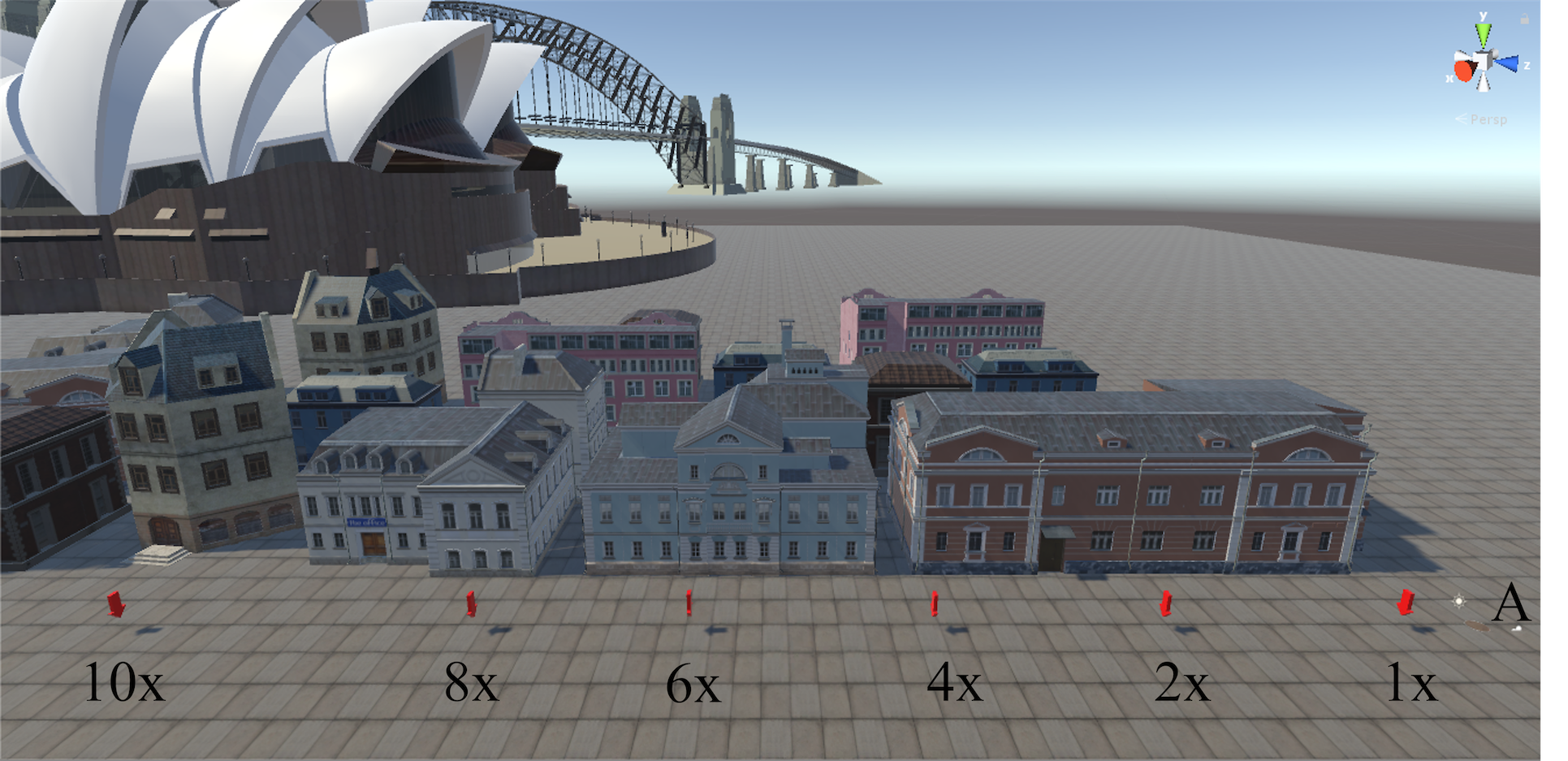}
    \caption{The side view of the virtual city. The red arrows indicate the mid-point of the virtual walking, where the participant turns and walks back to the starting point. Each red arrow corresponds to different levels of translation gain.}
    \label{fig:tg_fig}
\end{figure}

Our experiment used biomechanical measurement (full-body motion capture), neurological measurement (64-channel EEG), and self-reported questionnaires (Figure~\ref{fig:teaser}). 
During the experiment, the participant was instructed to walk toward a destination marked by a red arrow and back to the starting point inside a virtual city with increasing TG levels, including $1x, 2x, 4x, 6x, 8x, 10x$ (Figure \ref{fig:tg_fig}).
Based on the questionnaire responses, participants were divided into two groups: the \vrs group, i.e., the participants who experienced \newtext{strong} VR sickness \newtext{symptoms}, and the \novrs group, i.e., the participants with \newtext{little to }no VR sickness \newtext{symptoms}.
A between-group analysis was conducted to understand \toremove{how} \newtext{the relationship between} VR sickness \toremove{affects} \newtext{and} the users' biomechanics and neurological characteristics in the mobile VR setup.
Note that the experiment design does not randomize the TG level because the experiment was not intended to evaluate the effect of TG levels on VR sickness.

\subsection*{Contributions}

\begin{itemize}
\item To the best of our knowledge, this paper presents the first EEG-based study investigating VR sickness in a mobile VR setup, i.e., the user navigate a virtual environment with natural walking while wearing a VR headset. 
\item The experiment recorded both EEG and full-body motion capture data, \newtext{our preliminary results can} \toremove{which} contribute new evidence \toremove{and explanation} to the ongoing debate about the validity of postural instability theory \newtext{and inspire future work in this area}.
\item The related work section presents a detailed literature review of EEG-based experiments for VR sickness. The review summarises the experiment protocols and apparatus used in previous works and the observed brain dynamic changes associated with VR sickness. 
\end{itemize}

%% file: 20_related_work.tex
\section{Related Work} \label{prev_work}

The VR community has accumulated a large volume of research studying VR sickness, and we would like to refer readers to several excellent survey articles \cite{Rebenitsch2016, Saredakis2020, Chang2020-review} for complete reviews. 
This related work section complements these survey articles and focuses on works that used EEG to measure VR sickness and postural instability. 

\subsection{EEG and VR sickness}
Recent advances in BCI hardware and signal processing algorithms have enabled novel BCI applications and paradigms \cite{Abiri2019-EEG_review}. 
Passive BCI \cite{Zander2011-passive}, which continuously monitors the users' cognitive states, has been proven particularly valuable in many human-computer interaction scenarios. 
For example, Peck et al. \cite{Peck2013-fnirs_eval} used BCI to evaluate visual interfaces. Frey et al. \cite{Frey2016-framework} proposed an EEG-based framework that estimated the user's mental workload and attention from EEG signals to evaluate user experience. Yuksel et al. \cite{Yuksel2016-BACH} developed a BCI-based learning system that adaptively provides music learning tasks with different difficulty levels to enhance the user's learning efficiency.

There is a growing body of research on monitoring motion sickness and VR sickness from the VR community using the EEG-based passive BCI approach. (Table~\ref{tab:EEG}). One of the major benefits of using passive BCI for monitoring VR sickness is its capability to measure the sickness level continuously. Most VR works use questionnaires, such as Simulator Sickness Questionnaire (SSQ) \cite{Kennedy1993}, Virtual Reality Sickness Questionnaire (VRSQ) \cite{Kim2018-VRSQ}, or Cybersickness Questionnaire (CSQ) \cite{Stone2017-CSQ}, to measure the VR sickness symptoms during VR uses. Despite its many advantages \cite{2021_Hirzle_SSQ}\toremove{. T} \newtext{, t}he natural constraint of questionnaires is that they can only be conducted when the user is not actively performing the task. Some previous works used an additional joystick \cite{Chen2010, Chuang2016, Krokos2022-ik} or a physical dial \cite{mchugh2019-dial} to allow frequent reports of the level of sickness. However, these additional input modalities still introduce different levels of distraction. In contrast, the passive BCI approach provides continuous measurements and does not incur additional workload or distract users from their current tasks. 

Table~\ref{tab:EEG} lists research works that investigated VR sickness using the passive BCI approach. Because these works used a wide range of experiment protocols to induce VR sickness, we sorted this list of works accordingly to their experiment designs and apparatus. The most common protocol for inducing VR sickness is to expose the participants to video content with strong motions, such as virtual navigation \cite{Min2004, Kim2005, Park2008, Naqvi2015-iy, Li2020} or virtual flythrough \cite{Lim2021, Jang2022-wm, Krokos2022-ik, Nurnberger2021-nz, Wu2020-N2}. The virtual navigation videos mostly involved the urban virtual environment with streets, roads, and sometimes obstacles. 
The virtual flythrough videos show a wider range of content, from flying to spacewalking, which exhibit stronger roll and pitch motions. 
The participants were mostly instructed to stay in a stationary sitting posture and passively watched the video content throughout the experiment session. 
Keeping the participant stationary is common in EEG-based experiments because it will reduce the potential noise in the EEG signals due to body movements.
The participants' passive watching of identical video contents also ensures they perceive similar visual stimuli, allowing more precise time-locked EEG analysis if required.

Several works extend the 2D video setup by incorporating a motion platform to provide visual and vestibular stimuli. A series of works \cite{Lin2007-iq, Chen2010, Lin2013} used an experimental setup that included a CAVE and a vehicle mounted on a 6-degree-of-freedom motion platform. During the experiment, the participants behaved as a passenger in the vehicle as the vehicle automatically drove through a winding tunnel.
The same group later updated the protocol to include an additional driver actively driving the vehicle through the tunnel \cite{Huang2021-ca}.
Recent work from Recenti et al. \cite{Recenti2021} simulates seasickness with a virtual environment of a small boat floating on the sea. The participants wore an HMD while standing on a motion platform that generated synchronized vibrations to the virtual wave. 

As shown in Table~\ref{tab:EEG}, most previous works reported the fluctuations of low-frequency band power when the participant experienced VR sickness.
However, some changes in the power spectrum and the locations of brain activities are contradictory among these works.
For example, among works that induced VR sickness with videos on a 2D display, some works \cite{Kim2005, Lim2021, Jang2022-wm} reported a low-frequency power increase while others \cite{Min2004, Park2008} reported a power decrease. 
% results of previous works that investigated VR sickness using EEG. 
% Even experiments using similar protocol and hardware sometimes contradicted with each other\cite{Chen2010,Huang2021-ca}. 
Considering the difference in experiment protocols, hardware (displays and EEG headsets), and EEG signal processing protocols, such inconsistency in frequency band power is unfortunately not uncommon in BCI research. 
As a result, recent work \cite{Wu2020-N2} explores the reactive BCI paradigm \cite{Zander2011-passive}, which uses more reliable brain dynamic features (e.g., event-related potentials). However, such brain features require strict time-locked events, which significantly limit the design of VR scenarios. 
There are also works \cite{Li2020, Lin2013} focusing on using a machine learning approach to train classifiers that can detect or predict the onset of the VR sickness while putting less emphasis on the changes of spectrum powers and the underlying brain dynamics. 

Complementing these previous works, our work investigates EEG changes of VR sickness while the participant actively navigates the virtual environment with natural walking.
As most modern VR headsets allow users to move freely in the physical space, we believe our work represents a meaningful contribution to the understanding of VR sickness. 
Note that the mobile VR setup inevitably introduces additional noise into the EEG signals \cite{Makeig2009-mobi} and requires additional post-processing of the EEG signals. We describe our EEG data processing pipeline in detail in Section~\ref{sec-EEG_processing}.  

\subsection{Postural Instability}
\toremove{We used an experimental protocol that induces VR sickness by gradually increasing the disparity between the movements in the virtual environment and the physical space .}
\newtext{We used an experimental protocol} \cite{Cortes2019_TG} \toremove{This protocol was} inspired by the research works on redirected walking, originally developed for users to explore large virtual environments inside a limited physical space \cite{Steinicke2010, Suma2012, Nilsson2018}.  % Sun2016, Langbehn2017
These works carefully control the amplification below the user's detection threshold \cite{steinicke2009_threshold,grechkin2016_threshold, Kruse2018}. 
In contrast, we intentionally increase the translation gain so that the users can perceive the disparity. 
The protocol hypothesizes that the disparity of movements between the virtual and physical worlds would require the user to adjust their gait strategies to maintain postural stability. 
As the disparity gradually increases in the experiment, the users would need to consistently change their posture to fight the postural instability. 
%Maybe remove Bailey 2022
\newtext{According to different studies on postural instability \cite{Stanney2020a, Bailey2022}} \toremove{According to the postural instability theory}, as the users remain unstable, the level of VR sickness would also increase. 

There is a significant amount of works and evidences both for and against the postural instability theory, and we would like to direct our readers to a recent review \cite{palmisano2020_postural}. 
% We will also revisit some relevant works in the discussion section. 
Here we would like to focus the review on works that use VR and BCI to measure and understand human balance control. 
The main applications of these research works are in the domain of fall prevention for elderly adults or performance enhancement in rehabilitation settings \cite{Wittenberg2017}. 

The seminal works from Slobounov et al. investigate brain functions and behaviour outcomes using experimental methods such as self-initiated postural sways \cite{Slobounov2005, Slobounov2008} and postural instability induced by rotating the virtual environment \cite{Slobounov2013, Slobounov2015-EEGVR}. 
These works agreed on theta power increases in the frontal areas during postural instability. 
Chang et al. \cite{Chang2016-posture} induced postural instability with a virtual bus scenario on a motion platform to compare the brain activity of elderly adults with low and high fall-risk potential. The result suggested a strong correlation between postural-related cortical regions for the low fall-risk group.
Peterson et al. \cite{Peterson2018-short_term_balance} found that transient visual perturbations, i.e., a short roll of the virtual environment while the user is walking on a treadmill, could boost short-term motor learning, supported by the increased theta power and decreased alpha power in parietal and occipital regions and better balance performance. Many previous works also correlate the EEG signals with biomechanical measurements, such as the centre of pressure with force plate \cite{Slobounov2013} or gait kinematics with motion tracking \cite{Peterson2018-short_term_balance}.

Building on the methodologies of these previous works, we contribute the knowledge of the brain dynamic related to \toremove{the postural and gaits} \newtext{posture and gait} when the user experiences VR sickness in a mobile VR setup.

%% file: 30_experiment.tex
\section{Experiment}\label{exp_design}
\subsection{Participants}
Twenty-one healthy adults (17 males and four females) participated in the experiment. The mean age was $25.73$, with a standard deviation of $3.6$. All participants gave written informed consent and were compensated for their participation. All participants had a normal or corrected-to-normal vision. Among all participants, 13 had experience with 3D computer games, nine had experience in VR, and 15 had previous experience with motion sickness.

\subsection{Physical Space and Virtual Environment}
The physical space of the experimental environment was 3 meters by 5 meters (Figure \ref{fig:teaser}). In each trial, the participant walked from point A to point B and back to point A. Points A were marked with black tape on the ground to ensure the participant stood at the proper initial position. The distance between A and B was 5 m. 

The virtual size of the entire virtual scene was 100 m by 100 m. 
The virtual walking took place on a straight street 70 meters long and 5 meters wide. 
Figure \ref{fig:tg_fig} shows the side view of the road, and Figure \ref{fig:teaser} shows the participant's first-person perspective while performing the walking task. 
We marked both ends with a virtual utility hole and a virtual red arrow.

\newtext{
The virtual environment was developed using the Unity 3D game engine, version 2019.
The virtual environment images were transmitted to the VR HMD via cable.
}

\subsection{Apparatus}
The user's full body motion was captured with the OptiTrack system with 12 Flex 13 cameras at the capture rate of 120 frames per second. 
We used the \textit{Baseline + Toe marker} skeleton template with 41 tracked markers in the Motive:Body software for all participants. 
The HMD used in the experiment was the Oculus VR CV1, which has a 110-degree field-of-view resolution of 1080 x 1200 pixels per eye. 
\newtext{The cable of the VR headset hung on a ceiling hook to avoid affecting the participants' movements.}
The OptiTrack Unity Plugin, which only supported Oculus VR CV1 at the time of the experiment, was used to synchronize the data between the motion caption system and the virtual environment.  

The EEG data were recorded from 64 Ag/AgCI slim active sensors positioned according to the extended 10-20 electrode placement system. 
A portable and wireless LiveAmp system (Brain Products GmBH, Munich, Germany) was used to minimize the interference of natural walking in the virtual environment.
\newtext{The wireless LiveAmp receiver was on a waist bag, as seen in Figure \ref{fig:teaser}.}
The EEG data were acquired at a sampling rate of 500 Hz. 
The channel locations were assigned based on standard location, distributed by the EEGLAB toolbox \cite{Delorme2004, Delorme2011}. 
Contact impedance was maintained below $5 k\Omega$ for the EEG recording.

\subsection{Experiment Design}
% The experiment used a within-subject design, where the sole independent variable was the level of TG. 
The experiment induces VR sickness by gradually increasing the level of translation gain. 
The experiment starts from TG level $1x$, where the user's movements in the real world are directly mapped to the virtual environment without amplification. 
In each trial, the participant walks from the utility hole to the mid-point indicated by a red arrow, then back to the utility hole in the virtual city (Figure~\ref{fig:teaser}).
The TG level increases every 5-trials (a $block$), in the order of $1x, 2x, 4x, 6x, 8x, 10x$.
The experiment consists of six blocks, each corresponding to one TG level, and there are 30 trials for each participant. 
The participant walks the same physical distance in the real world in each trial, regardless of the TG level. The TG level augments the virtual walking distance and thus the different mid-point locations, indicated by the red arrows (Figure~\ref{fig:teaser}).

An informal pilot test led to our decision of using $10x$ gain as the upper bound of the TG level. 
The VR engineer in our team, who has extensive experience in different redirected walking tests, considered TG levels larger than $10x$ impractical and too difficult for the participant to complete even one trial. 
Our implementation applied motion magnification to all axes, which is different from the traditional implementation \cite{Interrante2007-TG_Navigate}, which amplifies the motion in the forward direction only. 
The decision was intended to ensure that participants were under constant motion magnification and postural instability throughout the experiment. 

Participants were split into the \vrs and \novrs groups based on the SSQ after the experiment. 
The \vrs group consists of participants who experienced \newtext{severe} VR sickness symptoms while interacting with the virtual environment.
The \novrs group consists of the participants that \toremove{did not} experience \newtext{minor or null} VR sickness \newtext{symptoms} during the experiment. 
Each group had two female participants.

Note that our experimental design does not randomize the order of TG levels as we do not intend to investigate the relationship between the TG level and the level of VR sickness. 
Instead, similar to previous EEG experiments that used multi-block designs \cite{Min2004, Park2008, Nurnberger2021-nz}, we compare the biomechanical and EEG data between \vrs and \novrs groups in each block to understand the effect of VR sickness \newtext{at different experimental stages}.
\newtext{
We want our outcomes to be the preliminary results of one of the first experiments to address VR sickness on a mobile VR setup.
}

The supplemental video shows two trials with the TG levels of $1x$ and $2x$. 
Please note that our participants did not agree to be recorded during the experiment. 
Thus the video is demonstrated by a researcher. 
\toremove{During the real experiment, the cable of the VR headset hung on a ceiling hook to avoid affecting the participants' movements.}

%% SECTION PROCEDURE
\subsection{Procedure}

\begin{figure*}[th]
    \centering
    \includegraphics[width=\textwidth]{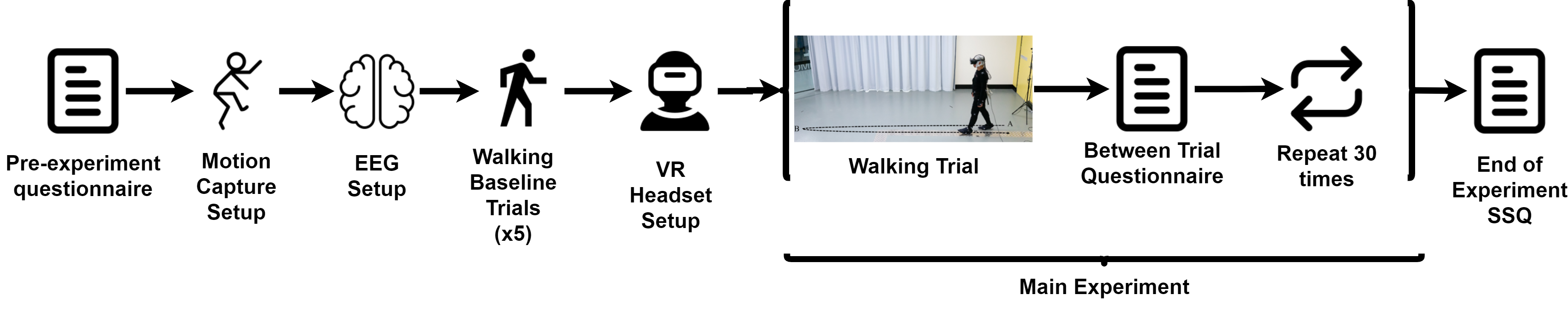}
    \caption{Flow of the experiment session.}
    \label{fig:exp_flow}
\end{figure*}

Figure \ref{fig:exp_flow} shows the complete experiment procedure. 
The session started with a \textit{pre-experiment questionnaire \newtext{(PEQ)}.} 
The \toremove{questionnaire} \newtext{PEQ} inquired about the participants' familiarity with 3D and VR technologies and their subjective assessment of their susceptibility to different types of motion sickness.
After completing the \toremove{questionnaire} \newtext{PEQ}, the researcher helped the participant put on the motion capture suit and ensured the participant's full-body motion within the experiment area could be successfully captured and displayed in the software. 
After completing the motion capture setup, the researcher proceeded with the EEG headset setup. The researcher injected conductive gels into each of the 64 electrodes and ensured its impedance shown on the EEG capture software was below $5 k\Omega$. 
The EEG signals of the closed-eye resting and eyes blinking were also visually inspected after the setup.  

After the EEG and motion capture systems were ready, the participant performed the baseline walking trials (without wearing the VR headset). 
The participant was instructed to walk between the two ends of the walking area, i.e., from position A to B, then back to A in Figure ~\ref{fig:teaser}. 
After the baseline walking, the researcher helped the participant put on the VR headset and adjusted the interpupillary distance. 
Wearing a VR headset would inevitably cause physical contact with the electrodes. Thus, re-calibration and re-gel were usually required to correct the position and impedance of the EEG electrodes. 
The setup was completed at this point, and the setup process took around 40 to 50 minutes on average.
% the \textit{VR Headset Setup} stage started. First, our participants wore the VR headset. Because recent studies have pointed out that the setup of the headset influenced the level of perceived sickness in different gender groups \cite{Munafo2017, Stanney2020}, we asked all participants to adjust the headset the first time they wore it. We intentionally adjusted the interpupillary distance to a random value between sessions, so our participants were forced to adjust it when first wearing the VR headset. After the user reported that she felt comfortable wearing the headset, she would spend some time getting used to the virtual environment without any level of TG. Finally, they received instructions on the task they needed to perform.

As stated in the experiment design section above, the main experiment consists of 6 blocks, each with 5 trials of one TG level. 
The TG level increases from $1x$ in the first block to $10x$ in the last block. 
In each trial, the participant walks toward a red arrow and back to the utility hole (Figure~\ref{fig:tg_fig}) in the virtual city. 
Once the participant reaches the utility hole, a virtual message notifies the completion of the trial. 
It then prompts the participant to report their level of sickness from 1 to 10, where ten corresponds to severe VR sickness symptoms.

%% file: 40_measurements.tex
\section{Measurements}\label{measurements_section}
\subsection{Questionnaires}
\newtext{Besides the PEQ, t}\toremove{T}his experiment uses two types of questionnaires. The first questionnaire is the \textbf{Between-Trial Questionnaire.} At the end of each trial, each participant reports, on a scale from 1 to 10, their feelings of dizziness, discomfort, nausea, fatigue, headache, and eyestrain. We chose these symptoms following previous studies \cite{Lin2013, Wilson2018}, which also used a sub-set of symptoms in trials to shorten the report time. % Seay2002

The second questionnaire is the \textbf{Post-Experiment Questionnaire.} Upon completing the VR session, we asked each participant for a full SSQ, followed by a semi-structured post-hoc interview session where the researchers engaged with the participant and recorded their subjective experience during the experiment.

\subsection{EEG Data Processing}
\label{sec-EEG_processing}
\newtext{
The following subsections describe our methodologies to process and analyze the EEG data of our experiment.
Mobile VR with EEG experiments have well-established procedures to ensure that EEG data gets the highest signal-to-noise ratio.
To ensure this, the following sub-sections briefly describe the same procedures as Do et al. \cite{Do2021}.
The extensive work of these authors describes the methodologies of mobile experiments with VR and EEG.

Our work also follows the procedures from studies that describe how to remove gait-related artifacts \cite{Ehinger2014, Liu2022}. 
These procedures help minimize gait-related activity and do not affect any of the conclusions of this study.}
%Even if our study does not focus on the gamma band in the sensorimotor region, which is strongly related to gait-related movements \cite{Seeber2015}.}
All the EEG data analysis steps were performed in EEGLAB \cite{Delorme2004, Delorme2011} version 14.1.2 in Matlab (Mathworks Inc., Natick, Massachusetts, USA). Figure \ref{fig:eeg_proc} shows a summary of the EEG processing steps.

\subsubsection{EEG Pre-processing}
The basis of the EEG pre-processing methods are previous EEG studies that use ambulatory VR setups \cite{Do2021, Liu2022a}. The EEG data was first down-sampled, applied a band-pass filter, and cleaned. The cleaning process includes using the \textit{CleanLine} plug-in (v1.03), removing flat segments, and removing bad channels. Next, an independent component analysis (ICA) was used to extract the maxima independent source activation. The adaptive mixture independent component analysis (AMICA) \cite{Palmer2007} was used for this step. Finally, the eye components were removed. A more detailed description of the pre-processing step can be found in Appendix A.
% TODO: Removed Do2021b, add if neccesary in future
% Next, the data was epoched into 35 segments, each representing a trial. To calculate the epoch length, we decided to use the first three-quarters of a walking trial after the participant turned to start walking towards the red arrow. Because of the noise derived from the 180-degree turn, the study lost several data sets. Therefore, the first three-quarters were used to avoid any noise from the body turning. To calculate the exact time, we averaged the times of each trial of each condition from all the participants. The final epoch length was defined based on the shortest time interval of all the walking segments, which was 12.45 seconds.

% For the noise removal, any values outside the $mean \pm 3 std$ were identified as noise values and were removed. During this process, we realized that the first trial of the experiment for all participants fell into this category. Further analysis revealed that most participants interrupted the trial to ask questions, adjust the headset, or did other things that caused noise in the EEG signals. This process also led to the data removal of two participants.

\subsubsection{Data Grouping}
% The data was divided into four brain regions: Frontal (Fz, F1, F2, F3, F4, F5, F6, F7, F8), Central (Cz, C1, C2, C3, C4, C5, C6), Parietal (Pz, P1, P2, P3, P4, P5, P7, P8), and Occipital (Oz, O1, O2). Only the Frontal and Central brain regions data was used for this experiment. The rest of the data were discarded from the study.
%  data from their corresponding channels were averaged. Then, based on the participant's SSQ responses, the data were divided into two groups: the \textit{VRS} group and the \textit{NoVRS} group.

\newtext{For our analysis, including all the 64 channels of data was not useful because, in EEG channel-based studies, there will be redundant information since the individual EEG channel correlates with its neighbor channels. 
Thus, we picked up the representative channels which reflect the major brain regions in human brain experiments \cite{Hayashi2022}. These are the midline-frontal region (Fz), central (Cz), parietal (Pz) and temporal (T8).}\toremove{We used representative channels based on their spatial location to investigate brain dynamics at specific brain regions. More specifically, the channels Fz, Cz, Pz, and T8 were used to explore the brain dynamic at the frontal, central, parietal, and temporal regions, respectively.} Subsequently, based on the participant's SSQ responses, the data were divided into two groups: the \textit{VRS} group and the \textit{NoVRS} group.

\subsubsection{EEG Post-Processing}
% Power Spectral Density (PSD), a widely used methodology to make channel-based comparisons on EEG data \cite{Cohen2017}, was calculated. The \textit{spectopo.m} function from EEGLAB was used to generate the PSD data for each group. PSD also allows for the separation of data by frequency. Hence, the data was divided into five different frequency bands. These bands are delta (1Hz - 4Hz), theta (4Hz - 8Hz), alpha (8Hz - 13Hz), beta (13Hz - 30Hz), and gamma (30Hz - 80Hz).

% The EEG signals recorded during the walking trials without a VR headset were used as a baseline. Because the signals at this stage represent the user walking without any of the cortical effects that the VR headset produces, these signals could help highlight the changes produced by VR sickness and postural instability in participants. 

% The procedure to highlight the cortical changes was subtracting the data at the baseline trials from the data at each main experiment trial. We followed the normalization procedure of EEG data presented by Cohen \cite{Cohen2014} and Cohen\cite{Cohen2017}. This procedure consists of log-subtracting the Baseline data from the data at each trial. At the end of this stage, the data was in the decibel (dB) notation.

The walking EEG data segment was extracted from the continuous EEG data into 35 segments, each representing a trial. To calculate the epoch length, we decided to use the first three-quarters of a walking trial after the participant turned to start walking toward the red arrow. Because of the noise derived from the 180-degree turn, the study lost several data sets. Therefore, the first three-quarters were used to avoid any noise from the body turning. To calculate the exact time, we averaged the times of each trial of each condition from all the participants. The final epoch length was defined based on the shortest time interval of all the walking segments, which was \toremove{12.45} \newtext{8} seconds.

For the noise removal, any values outside the $mean \pm 3 std$ were identified as noise values and were removed. During this process, we realized that the first trial of the experiment for all participants fell into this category. Further analysis revealed that most participants interrupted the trial to ask questions, adjust the headset, or do other things that caused noise in the EEG signals. This process also led to the data removal of two participants.

Then, the Event-Related Spectral Perturbation (ERSP) was used to transform the time series EEG data into the time-frequency domain (refer to \cite{Do2021}). 
\newtext{In the ERSP time-frequency domain, the noise trials were removed if they were outside the $mean \pm 3 std$. 
This method has demonstrated the feasibility of mobile EEG data analysis in previous studies \cite{Do2021}. Subsequently, }
the ERSP \newtext{data} \toremove{signals} recorded during the walking trials without a VR headset were used as a baseline. 
\newtext{EEG analysis requires normalizing the data with an appropriate baseline that removes any individual artifacts introduced by the user’s cognitive state before the study starts \cite{Cohen2014}.
To deal with these artifacts and highlight the changes produced by VR sickness and postural instability in participants, the signals of the user walking without any of the cortical effects that the VR headset produces were selected as the baseline.
}
\toremove{Because the signals at this stage represent the user walking without any of the cortical effects that the VR headset produces, these signals could help highlight the changes produced by VR sickness and postural instability in participants. }
\newtext{Finally, the statistical test was performed to check the significant difference between blocks (concatenated all the participant data given their sickness group) and without VR headset using a permutation test (n=2000 permutations) with false discovery rate correlation (FDR, $\alpha$=0.05). The non-statistical difference values of the group ERSP data were removed, and the significant difference values were retained for the final visualization on the grand average of all participants, e.g., Figure \ref{fig:cz_ersp}, \ref{fig:fz_ersp}, and \ref{fig:pz_ersp}}.

% Finally, we investigate the power change of the EEG at at specific frequency, including delta (1Hz - 4Hz), theta (4Hz - 8Hz), alpha (8Hz - 13Hz), beta (13Hz - 30Hz), and gamma (30Hz - 50Hz).  

\subsection{Biomechanical Measurements}
\subsubsection{Center of Mass Calculation}
The displacement between the CoM measured during the baseline walking trials without the VR headset and the CoM measured during virtual walking at different trials was calculated to analyze the change in CoM. The methodologies described by Lafond et al. \cite{Lafond2004} were used to calculate CoM. Because the speed of movement of the CoM varied among the different trials for each participant, we followed the procedure by Adistambha et al. \cite{Adistambha2008}. They used Dynamic Time Warping (from MATLAB's Dynamic Time Warping function) to align the different CoM signals.

\subsubsection{Gait Parameters}
The gait parameters we calculated were step distance, cadence, and time to complete a trial. Step distance was calculated based on the motion of the markers on the participant's feet. The step information was extracted following the work by Hreljac and Marshall \cite{Hreljac2000}. The function \textit{findpeaks.m} from MATLAB, with the \textit{MinPeakProminence} parameter set to 0.01 to remove all the noise peaks, was used to detect the upwards and downwards movements of the feet. These calculations helped calculate the number of steps a user used in each trial. Finally, we used each user's number of steps per minute to calculate cadence.

%% file: 50_result.tex
\section{Results}\label{results_section}

% \tim{Following the argument of prolong exposure to instability, you probably should provide additional regression analysis between all measurements and exposure time.}

Excluding the experiment setup time, the main experiment took about 25 minutes per participant. Among the 21 participants, five participants were removed due to hardware malfunction or excessive noise in the EEG data. 
Out of the 16 remaining participants, three terminated earlier due to severe VR sickness completing 16, 20, and 22 trials and one completed 34 of the 35 trials.

The Shapiro-Wilk test was used for the behavioural measurements to test their normality. 
If the data followed a normal distribution, a Welch's t-test was used to compare the measurements between the two groups. 
Otherwise, a Kruskal-Wallis H-test was used for the comparison. 
For the EEG data, to avoid the Multiple Comparison Problem~\cite{Maris2007}, we used the EEGLAB's permutation test (n=2000, FDR-correction)~\cite{Maris2007, Files2016} to identify the significant difference between the \textit{VRS} group and the \textit{NoVRS} group.
\newtext{We used the unbalanced version of the respective tests to test experiment blocks with an unbalanced number of trials.}

% a different static analysis procedure to avoid the Multiple Comparison Problem (MCP), which is common when analyzing physiological signals \cite{Maris2007}. We used 

\subsection{Behavioral Results}
\begin{figure}
    \centering
    \includegraphics[width=\columnwidth]{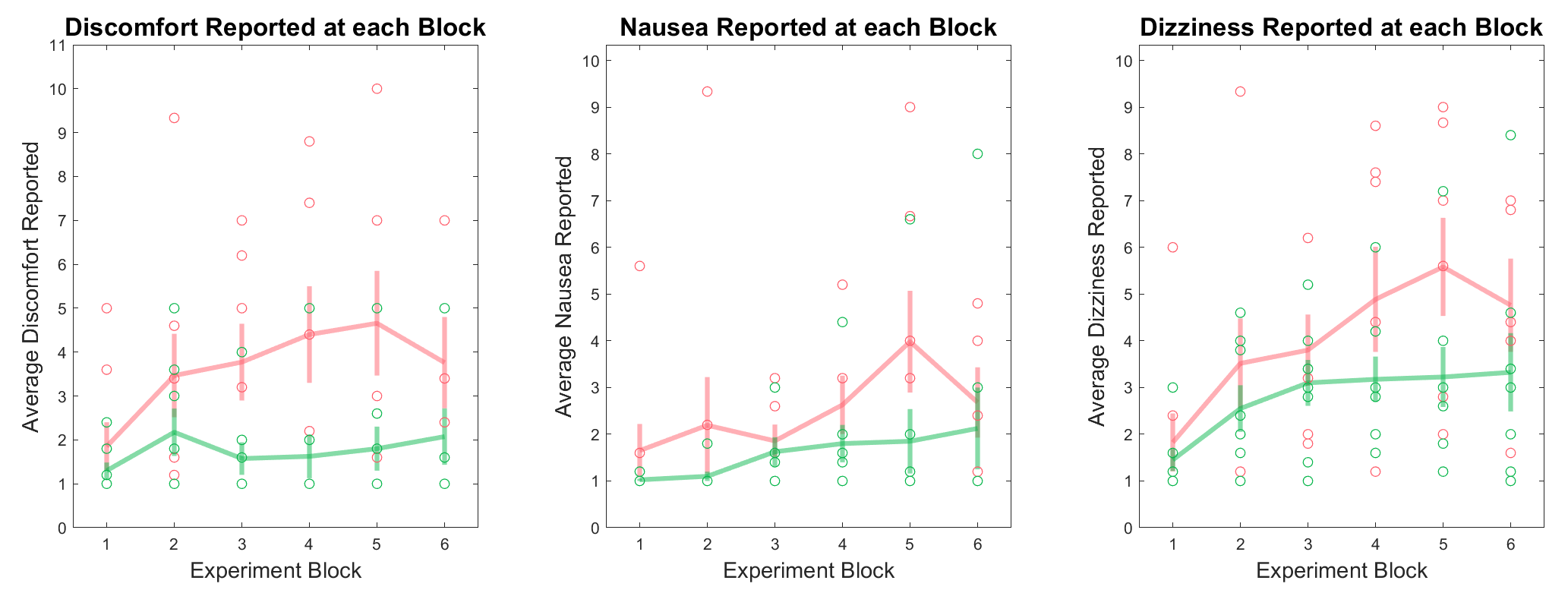}
    \caption{The discomfort, nausea and dizziness scores of the \textit{NoVRS} group (green) and the \textit{VRS} group (red). The bars represent the standard error. \newtext{The dots represent the individual average of each user, in each experiment block.}}
    \label{fig:questionnaire_results}
\end{figure}

\subsubsection{Post-Experiment SSQ}
Kennedy et al. \cite{Kennedy1993} and Stanney et al. \cite{Stanney1997} suggested a threshold of 18 on the total severity (TS) score as an indicator of a problematic level of sickness. 
The \textit{VRS} group consisted of 8 participants with a TS score of 18 or higher. 
The \textit{VRS} group had TS score of $\mu = 39.4850, \sigma = 17.378$, Nausea score of $\mu = 47.7, \sigma = 19.7497$, Occulomotor score of $\mu = 41.69, \sigma = 35.0886$ and Disorientation score of $\mu = 107.88, \sigma = 44.488$. 
\newtext{The \textit{NoVRS} group consisted of 8 participants with a TS score of 18 or lower.}
The \textit{\newtext{No}}\textit{VRS} group had TS score of $\mu = 6.0225, \sigma = 5.3411$, Nausea score of $\mu = 14.31, \sigma = 11.4025$, Occulomotor of $\mu = 9.475, \sigma = 13.2843$ and Disorientation score of $\mu = 12.18, \sigma = 15.6738$.

\subsubsection{Between-Trial Questionnaire}
\toremove{All participants reported symptoms of dizziness, discomfort, and nausea. 
Two participants reported changes in eyestrain, four reported increased fatigue, and three reported headaches.}
\newtext{Only three participants from the \novrs group did not suffer from discomfort and nausea. All participants (regardless of the group) suffered from dizziness. Six participants (3 from the \novrs group and three from the \vrs group) suffered from fatigue. Five participants suffered from headache (3 from the \vrs group and two from the \novrs group). Nine participants suffered from eyestrain (3 participants from the \vrs and 6 participants from the \novrs group). More information is located in the supplementary material}
% To visualize the between-trial questionnaire results, we averaged the responses from all symptoms at each trial. Those responses were then grouped per experiment block for each participant and averaged with the rest.
We performed statistical tests on the full scores of the between-trial questionnaire and the scores of each symptom between the \textit{VRS} group and the \textit{NoVRS} group. Statistical differences were found in the discomfort and nausea scores. The \textit{NoVRS} group had a lower discomfort score than the \textit{VRS} group in blocks 3, 4, and 5. The Nausea score of the \textit{VRS} was also significantly greater than that of the \textit{NoVRS} group in block 5 of the experiment. 

A summary of the descriptive statistics appears in Table \ref{tab:beh_results}. 
Figure \ref{fig:questionnaire_results} shows the discomfort and nausea scores.
\newtext{We only included these figures because they are the only measurements with significant statistical differences and where there is a visible difference between both groups.
We included the rest of the figures for the rest of the symptoms in our additional material.}

\input{51_table1}

\newtext{\subsection{Biomechanical Results}}
% \begin{figure}
%     \centering
%     \includegraphics[width=\columnwidth]{Figs/new_biomechanical.png}
%     \caption{Summary of the biomechanical measurements of the experiment. Green represents the \textit{NoVRS} group, and red represents the \textit{VRS} group. The bars represent the standard error.\tim{combine figure 6 and 7 for better layout.}}
%     \label{fig:biomechanical_results}
% \end{figure}

\begin{figure*}
    \centering
    \begin{subfigure}
        \centering
        \includegraphics[width=0.4\columnwidth]{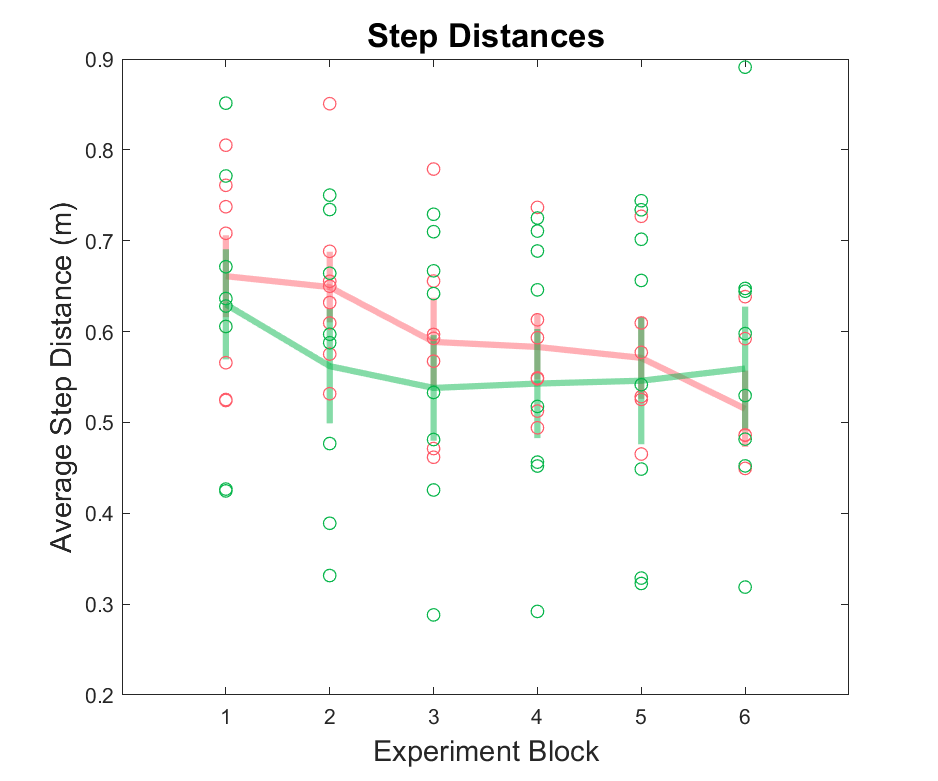}
    \end{subfigure}
    \begin{subfigure}
        \centering
        \includegraphics[width=0.4\columnwidth]{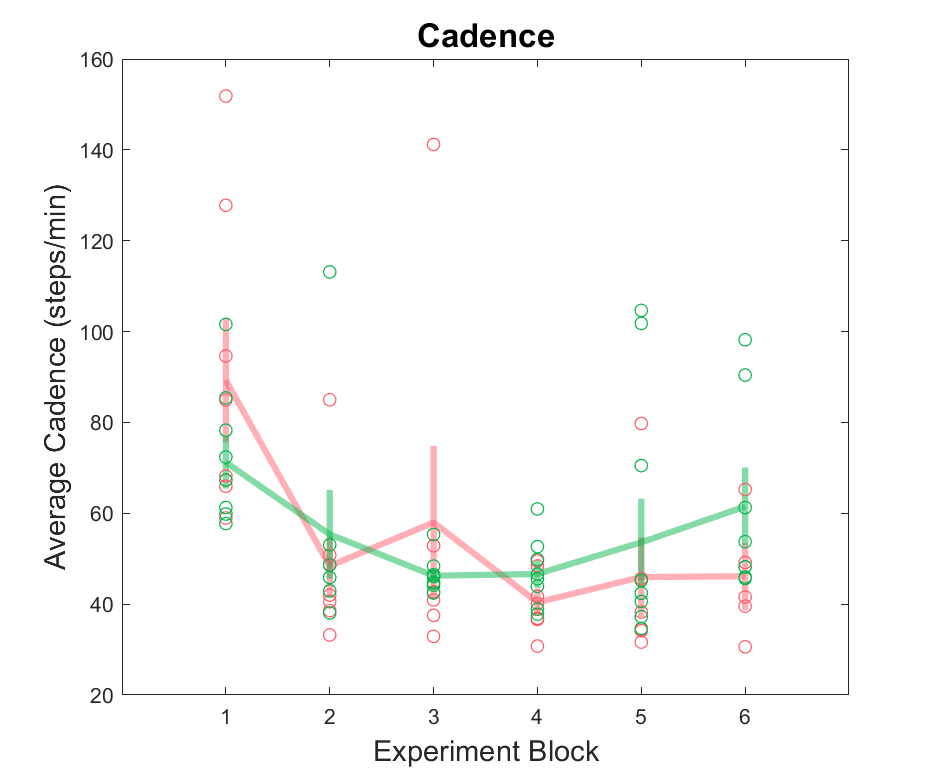}
    \end{subfigure}
    \begin{subfigure}
        \centering
        \includegraphics[width=0.4\columnwidth]{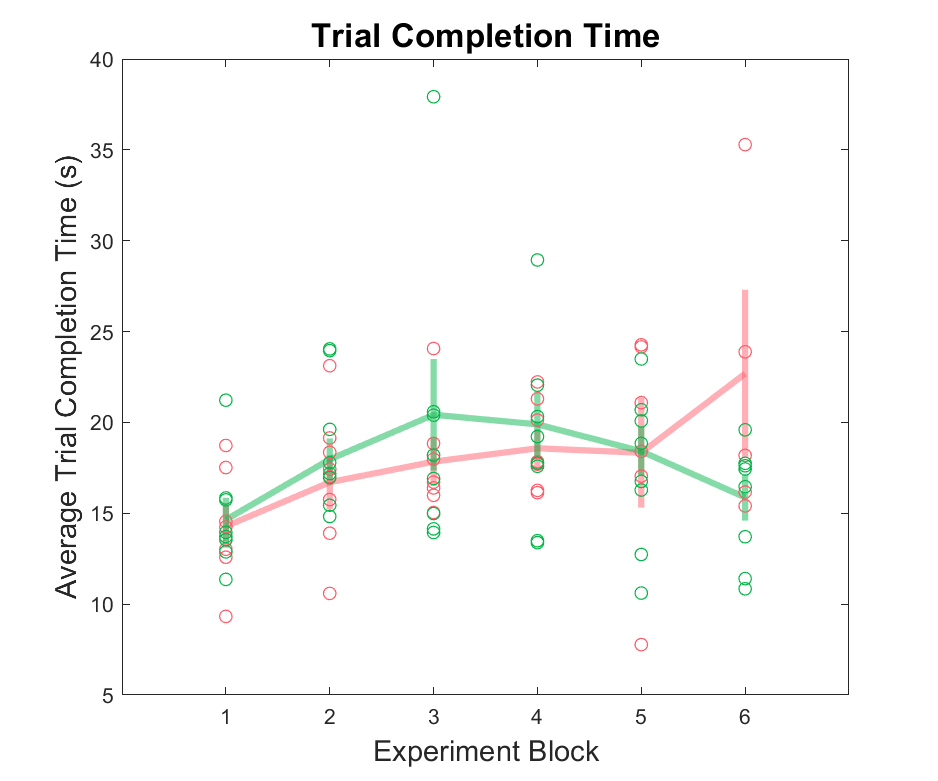}
    \end{subfigure}
    \begin{subfigure}
        \centering
        \includegraphics[width=0.4\columnwidth]{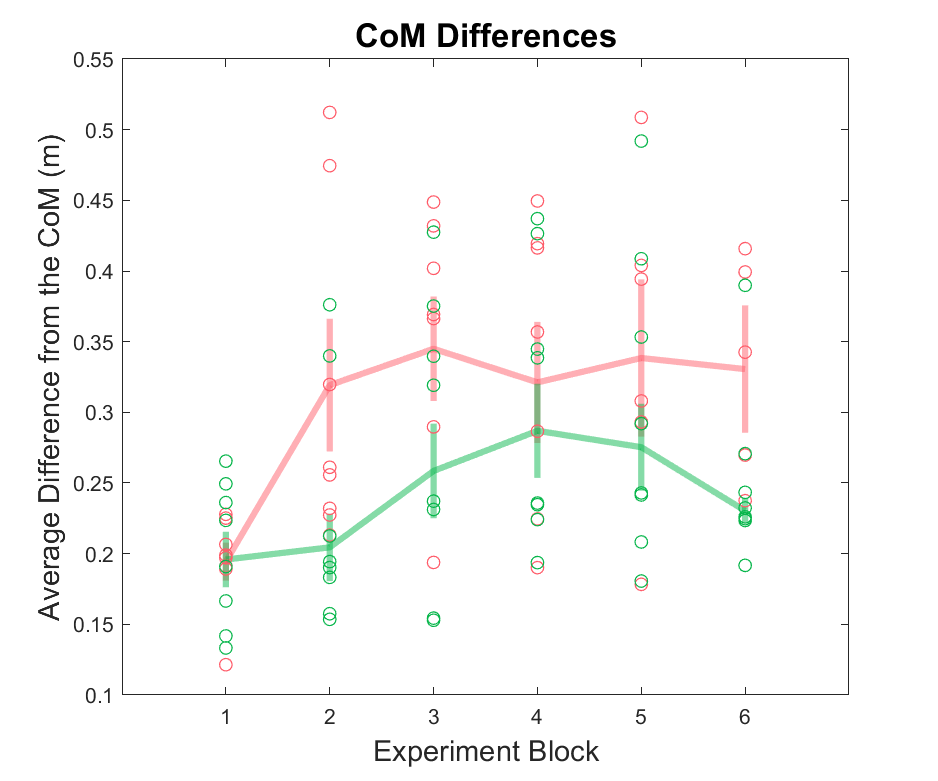}
    \end{subfigure}
    \begin{subfigure}
        \centering
        \includegraphics[width=0.4\columnwidth]{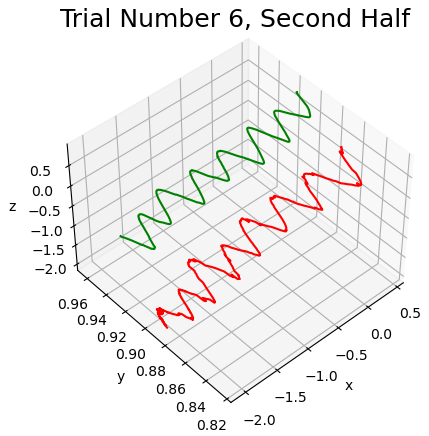}
    \end{subfigure}
    \caption{Biomechanical measurements of the \textit{NoVRS} group (green) and the \textit{VRS} group (red). From left to right: step distance, cadence, trial completion time, CoM differences, and an example 3D plot of CoM. We only show the second half of the trial to avoid visual clutter.}
    \label{fig:biomechanical_results}
\end{figure*}

% [width=.8\columnwidth]

Figure \ref{fig:biomechanical_results} shows the comparisons of biomechanical measurements between the two groups. 
Statistical differences were found on CoM in blocks 2 and 6 (see Table \ref{tab:beh_results}, last two rows).
At both blocks, the \textit{VRS} group showed a greater CoM displacement than the \textit{NoVRS} group. 
\toremove{None of the other measurements presented any other significant statistical difference.}
\newtext{No statistically significant differences were found in any of the other biomechanical measures.}

\input{54_table4.tex}

\subsection{EEG results}
\begin{figure*}[h]
    \centering
    \includegraphics[width=.8\textwidth]{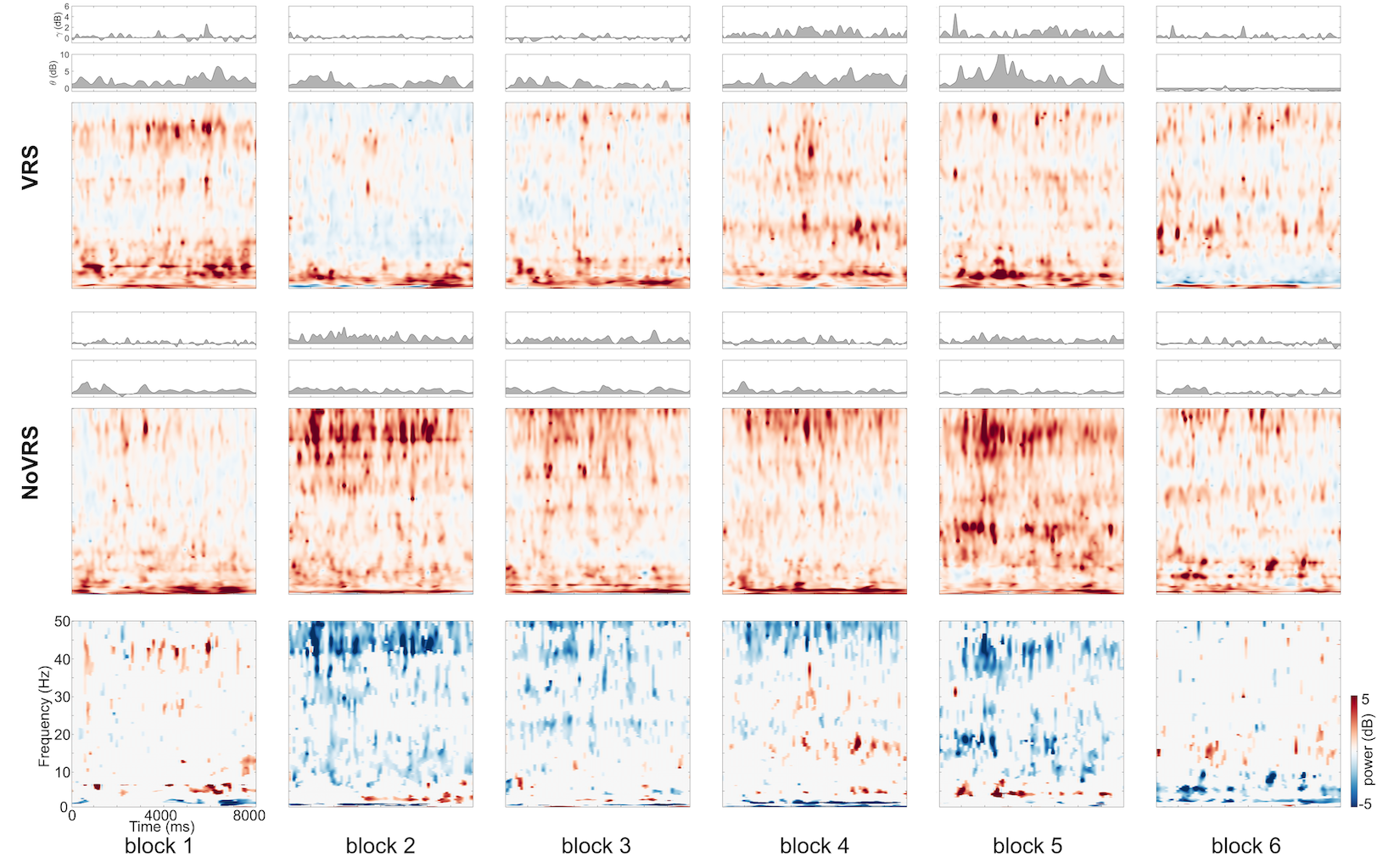}
    \caption{The Event-Related Spectral Dynamics (ERSP) of the \textbf{Fz} channel. The top subplot shows the ERSP of the \textit{VRS} group, the middle shows the ERSP of the \textit{NoVRS} group, and the bottom shows the statistical difference ($p < 0.05$) between the \textit{VRS} and \textit{NoVRS} groups.}
    \label{fig:fz_ersp}
\end{figure*}

Figures \ref{fig:fz_ersp}, \ref{fig:cz_ersp}, \ref{fig:pz_ersp}, and \ref{fig:t8_ersp} show the group ERSP results at the Fz, Cz, Pz and T8 channels respectively. 
Each figure contains the following information: the top rows show the ERSP of the \textit{VRS} group. 
The middle rows show the ERSP of the \textit{NoVRS} group. 
Finally, the bottom rows show the statistical difference between the two groups after the permutation test.

In the rows showing ERSP labelled with VRS and NoVRS, the red color represents power increases, while blue represents power decreases. 
In the last row, the red color denotes that the \textit{VRS} group has a statistically larger power than the \textit{NoVRS}. 
The blue color denotes that the \textit{VRS} group has statistically lower power.
A summary of the increases and decreases at each frequency appears in Table~\ref{tab:EEG}, at the bottom row.

Figure \ref{fig:fz_ersp} shows how the \textit{VRS} group had a significantly lower Fz power at the delta frequencies at blocks 1 through 4. The \vrs group had a significantly higher theta power at experiment blocks 1, 2 and 5. The alpha and beta frequencies were lower at experiment blocks 2 and 5, and the gamma frequencies were lower at experiment blocks 2, 3, 4, and 5.

Figure \ref{fig:cz_ersp} shows a significant decrease in Cz delta power through the experiment. A decrease in theta power in blocks 2, 3, 4, 5, and 6. A low-alpha power decrease from the \textit{VRS} group at experiment blocks 2, 3, 4, 5, and 6. A beta decrease at experiment blocks 5 and 6. And finally a significant decrease in the gamma power at experiment block 5. 

Figure \ref{fig:pz_ersp} shows a significant decrease in Pz delta power from the \textit{VRS} group at experiment blocks 1 and 3. The \vrs group also had significant theta decreases in experiment blocks 1, 3 and 6. Finally, there as a significant beta decrease in experiment block 2.

Finally, Figure \ref{fig:t8_ersp} shows a strong decrease in T8 delta frequency from the \textit{VRS} group throughout the experiment. There is also a significant decrease from the \textit{VRS} group at the theta frequencies at experiment blocks 1 through 3. There is an alpha frequency decrease at experiment blocks 2 and 3. A beta frequency decrease at experiment block 2, and a decrease at the gamma frequency in experiment blocks 2 and 5.

%% file: 51_table1.tex
\begin{table}[t]
    \centering
    \begin{tabular}{|c|c|c|c|c|}
        \hline
        \textbf{Measure} & \textbf{B} & \textbf{Stat}, \textbf{p-val} & $\mu$ & $\sigma$ \\
        \hline
        \hline
        Disc. & 3 & $\chi ^2(13)=4.92$, & \textit{VRS} 3.77 & 2.31 \\
         & & 0.02 & \textit{NoVRS} 1.57 & 1.04 \\
        \hline
        Disc. & 4 & $\chi ^2(13)=5.39$, & \textit{VRS} 4.4 & 2.91 \\
         & & 0.02 & \textit{NoVRS} 1.625 & 1.40 \\
        \hline
        Disc. & 5 & $\chi ^2(13)=4.40$, & \textit{VRS} 4.65 & 3.15 \\
         & & 0.03 & \textit{NoVRS} 1.8 & 1.41 \\
        \hline
        Nausea & 5 & $\chi ^2(13)=4.40$, & \textit{VRS} 3.98 & 2.88 \\
         & & 0.03 & \textit{NoVRS} 1.85 & 1.95 \\
        \hline
         \toremove{CoM} & \toremove{2} & \toremove{$\chi ^2(14)=4.41$,} & \toremove{\textit{VRS} 0.31} & \toremove{0.11} \\
         & & \toremove{0.03} & \toremove{\textit{NoVRS} 0.22} & \toremove{0.08} \\
        \hline
         \toremove{CoM} & \toremove{6} & \toremove{$\chi ^2(11)=4.2$,} & \toremove{\textit{VRS} 0.33} & \toremove{0.07} \\
         & & \toremove{0.04} & \toremove{\textit{NoVRS} 0.25} & \toremove{0.06} \\
        \hline
    \end{tabular}
    \caption{Behavioral measurements with significant differences. Disc. stands for Discomfort. The second column is the block number. The third column shows the statistic and p-value. The remaining two columns show the mean and the standard deviation.}
    \label{tab:beh_results}
\end{table}

% \begin{table}[t]
%     \centering
%     \begin{tabular}{|c|c|c|c|c|c|}
%         \hline
%         \textbf{Measurement} & \textbf{Block} & \textbf{Statistic} & \textbf{p-value} & $\mu$ & $\sigma$ \\
%         \hline
%         \hline
%          CoM & 2 & $\chi ^2(14) = 4.41176$ & 0.03569 & \textit{VRS} 0.31198 & 0.11688 \\
%         & & & & \textit{NoVRS} 0.22592 & 0.0843 \\
%         \hline
%         CoM & 6 & $\chi ^2(11) = 4.2$ & 0.04042 & \textit{VRS} 0.33298 & 0.07822 \\
%         & & & & \textit{NoVRS} 0.2503 & 0.06055 \\
%         \hline
%         Discomfort & 3 & $\chi ^2(13) = 4.92682$ & 0.02644 & \textit{VRS} 3.77143 & 2.31352 \\
%         & & & & \textit{NoVRS} 1.575 & 1.04983 \\
%         \hline
%         Discomfort & 4 & $\chi ^2(13) = 5.39343$ & 0.02021 & \textit{VRS} 4.4 & 2.91204 \\
%         & & & & \textit{NoVRS} 1.625 & 1.40789 \\
%         \hline
%         Discomfort & 5 & $\chi ^2(13) = 4.40859$ & 0.03576 & \textit{VRS} 4.65714 & 3.15534 \\
%         & & & & \textit{NoVRS} 1.8 & 1.41825 \\
%         \hline
%         Nausea & 5 & $\chi ^2(13) = 4.40859$ & 0.03576 & \textit{VRS} 3.98096 & 2.87949 \\
%         & & & & \textit{NoVRS} 1.85 & 1.95009 \\
%         \hline
        
%     \end{tabular}
%     \caption{Summary of the descriptive statistics and significant differences in the different behavioural measurements.}
%     \label{tab:beh_results}
% \end{table}

%% file: 54_table4.tex
\begin{table}[t]
    \centering
    \color{blue}
    \begin{tabular}{|c|c|c|c|c|}
        \hline
        \textbf{Measure} & \textbf{B} & \textbf{Stat}, \textbf{p-val} & $\mu$ & $\sigma$ \\
        \hline
        \hline
         CoM & 2 & $\chi ^2(14)=4.41$, & \textit{VRS} 0.31 & 0.11 \\
         & & 0.03 & \textit{NoVRS} 0.22 & 0.08 \\
        \hline
        CoM & 6 & $\chi ^2(11)=4.2$, & \textit{VRS} 0.33 & 0.07 \\
         & & 0.04 & \textit{NoVRS} 0.25 & 0.06 \\
        \hline
        
    \end{tabular}
    \caption{Biomechanical measurements with significant differences. CoM stands for Center of Mass. The second column is the block number. The third column shows the statistic and p-value. The remaining two columns show the mean and the standard deviation.}
    \label{tab:biom_results}
\end{table}

%% file: 60_discussion_v2.tex
\section{Discussion}\label{discussion_section}
\subsection{Posture Instability Theory and EEG}
\toremove{There were few statistical differences between \textit{VRS} and \textit{NoVRS} groups on gait parameters and biomechanical measurements. 
There was no statistical difference at the session level between the two groups. 
Only at the block level can we find significant statistical differences in CoM between two groups}
\newtext{No biomechanical measurements presented statistical differences but the difference in the Center of Mass. CoM showed differences} within blocks 2 and 6.
Block 2 has a corresponding TG level of $2x$ and was the first block when the participant experienced a perceivable disparity in body motion between real and virtual worlds. 
The result suggests that the \textit{VRS} group initially had trouble adapting to the new perceivable TG level. Indeed, during the post-hoc interview, multiple users in the \textit{VRS} group reported being surprised by the change in the TG level and commented that more time was needed to adapt to the sudden change in the virtual environment.
\toremove{
However, overall, the biomechanical result seems to concur with previous works that found no significant changes in the motion patterns when the user experienced VR sickness.}% \cite{ Akizuki2005, cobb1998_no_PIT, Howard2021}. 
% One possible explanation could be that the walking task was simple for the participants. Despite the participants in the \textit{VRS} group experiencing VR sickness symptoms, they could still maintain reasonable postural stability during most of the trials without significantly changing their gaits.

\newtext{
The statistically significant differences in the CoM appeared at Block 2, one experimental block before we found statistical differences in the BTQ.
These findings confirm the primary assumption of the postural instability theory, which states that postural differences will appear before the onset of subjective symptoms~\cite{Riccio1991:PIT}. 
Our EEG results in the central-theta frequency corroborate these findings.
Different studies with VR HMD have reported these changes in participants that struggle to adjust posture or to maintain upright balance \cite{Sipp2013, Hulsdunker2015, Edwards2018}. 
Given the relationship between the changes in posture and this signal, observation of the central-theta frequency can be used as an alternative method to identify if the participant is suffering from postural changes that lead to VR sickness.
}

\newtext{Our}\toremove{The} EEG \newtext{preliminary results} \toremove{data} yielded more important insights into the participants' cognitive performance. 
The EEG result suggests that \textit{VRS} group did maintain postural control but at the cost of a higher mental effort.
The \textit{VRS} group has a significantly lower alpha power through blocks 2, 3, and 4.
It has been well-established that alpha power (in the parietal brain region) was inversely correlated with mental workload \cite{klimesch1999_workload, Tao2019-workload}. % TODO: Candidate to remove is klimesch1999_workload
The suppression of alpha power was also reported when the participant consciously adjusted their posture while standing \cite{Slobounov2008, Wittenberg2017} and while walking \cite{Beurskens2016}. 
Interestingly, when walking with a prosthesis, a similar reduction in alpha and gamma bands have been reported when the cognitive-motor task demands increased \cite{shaw2019_limb,pruziner2019_limb}.
In addition, the increased theta power in the frontal area of the \textit{VRS} group also concurs with a series of works from Slobounov et al. \cite{Slobounov2005, Slobounov2008, Slobounov2013, Slobounov2015-EEGVR} which found the theta power increased when the participant experienced postural sways. 
% It is another evidence that \textit{VRS} group experienced postural instability during the experiment despite there was no significant difference in the biomechanical measurements. 
In sum, the lower alpha power (in frontal and parietal) and higher theta power (in frontal) of the \textit{VRS} group seem to suggest that although \textit{VRS} group could complete the tasks without significant changes in postural stability or gait parameters, they still found the task mentally challenging, which took more effort to complete them.

For the \textit{NoVRS} group, their higher delta power also supports the argument that postural instability theory plays an important role in the mobile VR setup. 
The increased delta and theta powers were found when the participant was capable of engaging in cognitive-motor tasks, such as visual information processing and postural adjustment \cite{Harada2009, Kline2014}. 
Presacco et al. \cite{Presacco2011} also observed an increased delta power when the participants engaged in precision walking tasks.
These findings suggest that the \textit{NoVRS} group exhibited a better cognitive capability to continuously update their postural control strategy when compared to the \textit{VRS} group.

In sum, despite both \textit{VRS} and \textit{NoVRS} groups having similar biomechanical measurements, the EEG data provides additional insights into the cognitive performance differences. The brain activities of \textit{VRS} group exhibited similar patterns that were previously linked to higher cognitive workload and postural instability. In contrast, the \textit{NoVRS} group showed brain activities that were expected when the participant is capable of performing complex motor tasks such as postural balance control and fine walking movements. 
% In contrast, the lack of these activities in \textit{VRS} group could indicate their reduced cognitive performance because of the VR sickness.

\subsection{Future VR Headset Design}

Table~\ref{tab:EEG} shows a list of works that used EEG to measure and understand VR sickness. Our experiment used a novel experimental protocol in a mobile VR setup and resulted in some contradictory EEG results from some previous works. As discussed in the previous section, the difference might be attributed to the cognitive activities related to postural control. In contrast, the EEG difference in previous works might represent the brain activities related to the sensory conflict. 
Nevertheless, most works, including ours, seem to agree that the fluctuation of low-frequency power along the frontal and parietal midline correlates with VR sickness. 

This finding suggests that the ideal locations for an EEG add-on for VR sickness detection on a VR headset might be at the frontal and parietal midline region. Because the changes in the low frequency seem to be a common phenomenon in most VR sickness experiments, it might be possible to train a single classifier that can be generalised to different virtual environments and VR setups (stationary or walking).

The result also suggests that postural instability theory is worth further investigation as most consumer VR headsets allow natural walking in the virtual environment. In addition, the biomechanical measurements alone might not be able to capture the full picture of the postural instability theory. Additional measurements such as EEG or other biosignals might be necessary for the future investigation alone this direction.

%% file: 70_conclusion.tex
\section{Limitations and Future Work}\label{limits_future}
One of the main limitations of this study is the number of participants. Due to the pandemic constraint, IRB in our university does not allow prolonged contact with participants. 
As a result, we could only recruit 21 participants for the experiment. 
More participants will strengthen the result and reduce the risk of having a biased group. 

Previous VR sickness works have suggested that participants experiencing VR sickness also reported higher task load~\cite{Chang2020-review}. Recent works also found that the task workload might also affect VR sickness~\cite{sepich2022_workload}. One of our EEG-based findings is that the participant with VR sickness could maintain postural stability at the cost of a higher mental workload. We did not use the NASA-TLX task load questionnaire at the end of the experiment because we originally considered the walking task simple enough for all participants. Retrospectively, we believe a between-trial workload questionnaire might provides additional supports our EEG findings. However, the questionnaire might not be able to distinguish whether the workload is because of VR sickness or the effort to maintain postural control.
% However, we did not use the NASA-TLX workload questionnaire at the end of the experiment because we suspect 
% Our work focused on EEG and postural instability and found interesting correlation between 

In this experiment, we split the participant into two groups based on the SSQ questionnaire score, similar to previous EEG experiments \cite{Jang2022-wm, Lim2021}. However, as mentioned in the recent work by Hirzle et al. \cite{2021_Hirzle_SSQ}, many important factors related to VR sickness, such as digital eye strain and ergonomics, are largely ignored by SSQ. In particular, the ergonomic symptoms, such as anxiety about the headset's weight and how the headset affects the movement, are relevant to the postural instability theory. It would be interesting to group participants based on the more comprehensive VR sickness questionnaire and re-examine the result.
% Firstly, a considerable number of participants could not finish the experiment. Having a similar or equal number of participants at the end of the experiment would have helped to understand more about their status. More participants would have strengthened the EEG analysis and helped us gain more insights from the behavioral data. Secondly, the global pandemic prevented us from collecting more data to make up for those participants who could not finish the experiment.

% Using the SSQ questionnaire allowed us to divide our participants into two groups. However, we previously considered using the between-trial questionnaire to group the data. However, several problems and limitations arose when doing this analysis. We encountered a heavily unbalanced number of trials assigned to different scores. For example, at some stage of analysis, we categorized trials with a score of 10 and below as having no sickness. Anything above 40 points was categorized as high sickness. The values in between were labeled as a medium level of sickness. We discovered that 70 \% of the trials had a score of medium or no sickness. Furthermore, only 1.4 \% of the trials scored above 40 points. If any future works decide to study the level of sickness per trial, a simple system (similar to that of \cite{Chen2010}) would be more helpful than a questionnaire with different symptoms and a large range of values.

\newtext{
Our work contains several limitations in the area of data analysis.
The current analysis of CoM focused on the spatial dynamics of postural instability.
We envision studying the temporal dynamics of the CoM of participants to provide a superior analysis on the status of the CoM and its relationship with the EEG signals.

A temporal analysis of the EEG results can shed some more information on the nature of our results. 
Our results are applicable considering that the level of VR sickness will gradually increase, which only applies to a certain number of VR applications. 
Comparative analysis within the group between each experiment block can help us understand more about the additive effects of VR sickness. 
Furthermore, another experiment randomizing the levels of TG can help understand the differences between additive effects vs. spontaneous changes of TG
}

The current work analyzed the EEG data using a channel-based approach and reported the EEG result along the midline. 
The 64-channel EEG signals are mainly used for noise removal in the pre-processing stage. 
With 64-channel EEG data, it might also be possible to localize the electrical activity in the brain using approaches such as sLORETA~\cite{pascual2002_loreta}. 
We have made some initial attempts in this direction, but the mobile VR setup and potentially the VR sickness symptoms significantly affect the quality of the results. 
Further investigation is needed to localize the distributed cortical sources of EEG activity.

% localizing the electrical activity in the brain
% the component-based analysis that could better locate the 
% explore the brain's cortical activity in the independent brain component manner. Nevertheless, the current results in EEG-channel-based analysis would help us point out the most noticeable influences in later ICA analysis, which demonstrated the power in a similar ambulatory EEG setup \cite{Do2021}.

\section{Conclusion}\label{conclusions}
This paper presents the first EEG-based study on VR sickness when the user actively walks in a virtual environment while wearing a VR headset. 
A comprehensive list of EEG-based studies on VR sickness was presented in the paper for guiding the future research.
Our study used multiple measurements, including full-body motion capture and EEG. 
\toremove{Although there was no significant differences in biomechanical measurements, the} \newtext{Our} EEG result\newtext{s together with the results in the difference of Center of Mass} suggests that the \textit{VRS} group spent higher cognitive efforts to maintain the postural balance during the experiment.
% shows significant brain features that link to fine postural control between \textit{VRS} and \textit{NoVRS} groups. 
The result suggests that the postural instability theory might play an important role in understanding VR sickness in mobile VR setups.

% In this paper, an ambulatory VR experiment involving motion capture for Center of Mass measurement and gait kinematics, as well as EEG signals, was implemented to study the effects of ambulatory VR on the level of VR sickness and postural instability of VR users. Our participants were divided into the \textit{VRS} group, i.e., the group that suffered from VR Sickness, and the \textit{NoVRS} group, i.e., the group that did not report significant changes in their symptoms of VR sickness. 

% Our study demonstrates how EEG helps understand the user's cognitive state and why they suffer VR sickness in a scenario. Comparing our study to previous EEG and VR sickness studies, we concluded that the delta frequency should not be used as an indicator of VR sickness, at least in the context of mobile VR. Moreover, our study has identified how the changes in the alpha frequency can be used as predictors of struggle to interact with a mobile, virtual environment.

%% file: 71_EEG_Table.tex
% \begin{tabular}{|l|p{0.1\linewidth}|p{0.1\linewidth}|p{0.1\linewidth}|p{0.1\linewidth}|p{0.1\linewidth}|p{0.1\linewidth}|p{0.1\linewidth}|p{0.1\linewidth}|p{0.1\linewidth}|p{0.1\linewidth}|}
% \begin{tabular}{@{}llllm{0.1\textwidth}m{0.1\textwidth}m{0.1\textwidth}m{0.1\textwidth}m{0.1\textwidth}m{0.1\textwidth}l@{}}

\newcolumntype{P}[1]{>{\RaggedRight\hspace{0pt}}p{#1}}

\afterpage{%
\begin{landscape}% Landscape page
\centering % Center table
\begin{table}[]
\begin{tabular}{@{}lP{0.1\textwidth}P{0.1\textwidth}P{0.1\textwidth}P{0.1\textwidth}P{0.1\textwidth}P{0.08\textwidth}P{0.2\textwidth}P{0.1\textwidth}P{0.5\textwidth}l@{}}
\toprule
\textbf{Paper} & \textbf{Frontal} & \textbf{Central} & \textbf{Parietal} & \textbf{Occipital} & \textbf{Temporal} & \textbf{Protocol} & \textbf{Experiment} & \textbf{Apparatus} & \textbf{EEG} \\ \midrule
Chelen 1993\cite{Chelen1993-xj} & $\delta \uparrow$  $\theta \uparrow$ &  &  &  & $\delta \uparrow$ $\theta \uparrow$ & Passive & Coriolis stimulation & Coriolis stimulation & 14-ch \\
Min 2004\cite{Min2004} & $\theta \downarrow$ & $\theta \downarrow$ &  &  &  &  Passive & Virtual driving video & 2D display & 2-ch (Fz Cz) \\
Li 2020 \cite{Li2020}$^{\ast\ast}$& &  &  &  &  & Passive & Virtual navigation video & 2D display & 8-ch \\
Lim 2021 \cite{Lim2021}& $\delta \uparrow$ $\theta \uparrow$ $\alpha \downarrow$ & $\delta \uparrow$ $\theta \uparrow$ $\alpha \downarrow$ &  &  &  & Passive & Virtual fly video & 2D display & 256-ch \\
Jang 2022 \cite{Jang2022-wm}& $\delta \uparrow$ $\alpha \downarrow$ &  & $\delta \uparrow$ $\alpha \downarrow$ & $\delta \uparrow$ $\alpha \downarrow$ & $\delta \uparrow$ $\alpha \downarrow$ & Passive & Virtual fly video & 2D display & 256-ch \\
Naqvi 2015 \cite{Naqvi2015-iy}& &$\theta \downarrow^\ast$  &  &  &  & Passive & Virtual navigation video & 2D and 3D displays & 128-ch \\
Kim 2005\cite{Kim2005} & $\delta \uparrow$ $\beta \downarrow$ &  & $\beta \downarrow$ &  & $\delta \uparrow$ $\beta \downarrow$ & Passive & Virtual navigation video & CAVE & 9-ch \\
Park 2008 \cite{Park2008} & $\theta \downarrow$ & $\theta \downarrow$ &  &  &  & Passive & Virtual navigation video & CAVE & 2-ch (Fz Cz) \\
Lin 2007 \cite{Lin2007-iq}& $\alpha \uparrow$ $\beta \uparrow$ &  & $\alpha \uparrow$ $\beta \uparrow$ &  &  & Passive & Autopilot on a winding road & CAVE + 6 DoF driving simulator & 32-ch \\
Chen 2010 \cite{Chen2010} & $\alpha \downarrow$ &  & $\alpha \downarrow$ & $\delta \uparrow$ $\theta \uparrow$ &  & Passive & Autopilot on a winding road & CAVE + 6 DoF driving simulator & 32-ch \\
Chuang 2016 \cite{Chuang2016}& $\alpha \uparrow$ $\gamma \uparrow$ &  & $\alpha \uparrow$ $\gamma \uparrow$ & $\alpha \uparrow$ $\gamma \uparrow$ &  & Passive & Autopilot on a winding road & CAVE + 6 DoF driving simulator & 32-ch \\
Huang 2021 \cite{Huang2021-ca}& $\alpha \uparrow$ &  & $\alpha \uparrow$ & $\alpha \uparrow$ &  & Passive \& Active & A driver actively drive on a winding road with a passenger & CAVE + 6 DoF driving simulator & 64-ch x 2 \\
Recenti 2021 \cite{Recenti2021}& & $\beta \downarrow^\ast$ &  & & & Passive & Simulation of standing on a floating small boat & HMD + moving platform & 64-ch (dry)\\

Kim 2019 \cite{Kim2019}&  &  &  & $\alpha \uparrow$ &  & Passive & Videos of roller coaster ride diving driving airplane control & HMD & 62-ch \\
Wu 2020 \cite{Wu2020-N2}$^{\ast\ast}$& $N2\uparrow$ $P3\downarrow$ & $N2\uparrow$ $P3\downarrow$ & $N2\uparrow$ $P3\downarrow$ & &  & Passive & Virtual navigation video & HMD & 24-ch \\
Nurnberger 2021 \cite{Nurnberger2021-nz}&  & $\delta \uparrow$ $\theta \uparrow$ $\alpha \uparrow$ & $\delta \uparrow$ $\theta \uparrow$ $\alpha \uparrow$ & $\delta \uparrow$ $\theta \uparrow$ $\alpha \uparrow$ & $\delta \uparrow$ $\theta \uparrow$ $\alpha \uparrow$ & Passive & Virtual fly controlled by an observer & HMD & 18-ch \\
Krokos 2022 \cite{Krokos2022-ik}&  & $\delta \uparrow$ $\theta \uparrow$ $\alpha \uparrow$ &  &  &  & Passive & Virtual fly video& HMD & 14-ch (dry) \\

&&&&&&&&&&\\
\textbf{Ours} & $\delta \downarrow$ $\theta \uparrow$ $\alpha \downarrow$ $\beta \downarrow$ $\gamma \downarrow$ & $\delta \downarrow$ $\theta \downarrow$ $\alpha \downarrow$ $\beta \downarrow$& $\delta \downarrow$ $\theta \downarrow$ $\beta \downarrow$ &  & $\delta \downarrow$ $\theta \downarrow$ $\beta \downarrow$ $\gamma \downarrow$& Active & Virtual navigation with natural walking & HMD & 64-ch \\ \bottomrule
\end{tabular}
\caption{Previous EEG-based works on motion sickness and VR sickness. The table is sorted according to the apparatus and experiment protocol used. All experiments except ours require the participants to stay in a sitting posture. All but one previous experiment \cite{Huang2021-ca} require the participants to perceive the pre-generated visual stimuli passively.  *\cite{Naqvi2015-iy, Recenti2021} report the average power changes across all chs.**\cite{Li2020} did not report frequency band activities but only focused on the classification of the signal. \cite{Wu2020-N2} used the amplitude of event-related potential (ERP, N2 and P3) as the indicators for VR sickness. 
}
\label{tab:EEG}
\end{table}

\end{landscape}% Landscape page
}%after page

%% file: 72_EEG_Pipeline.tex
\begin{figure*}[tb]
    \centering
    \includegraphics[width=\textwidth]{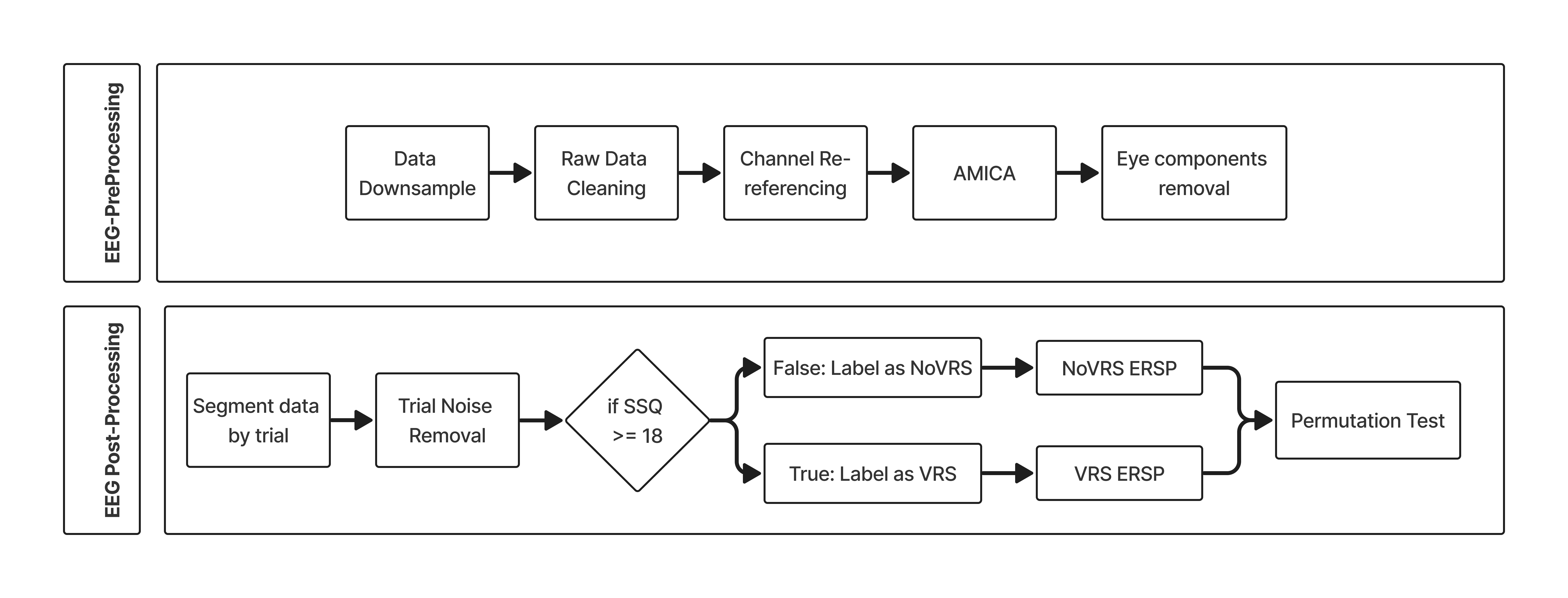}
    \caption{EEG processing pipeline summary.}
    \label{fig:eeg_proc}
\end{figure*}

%% file: 73_EEG_ERSP.tex
\begin{figure*}[h]
    \centering
    \includegraphics[width=.8\textwidth]{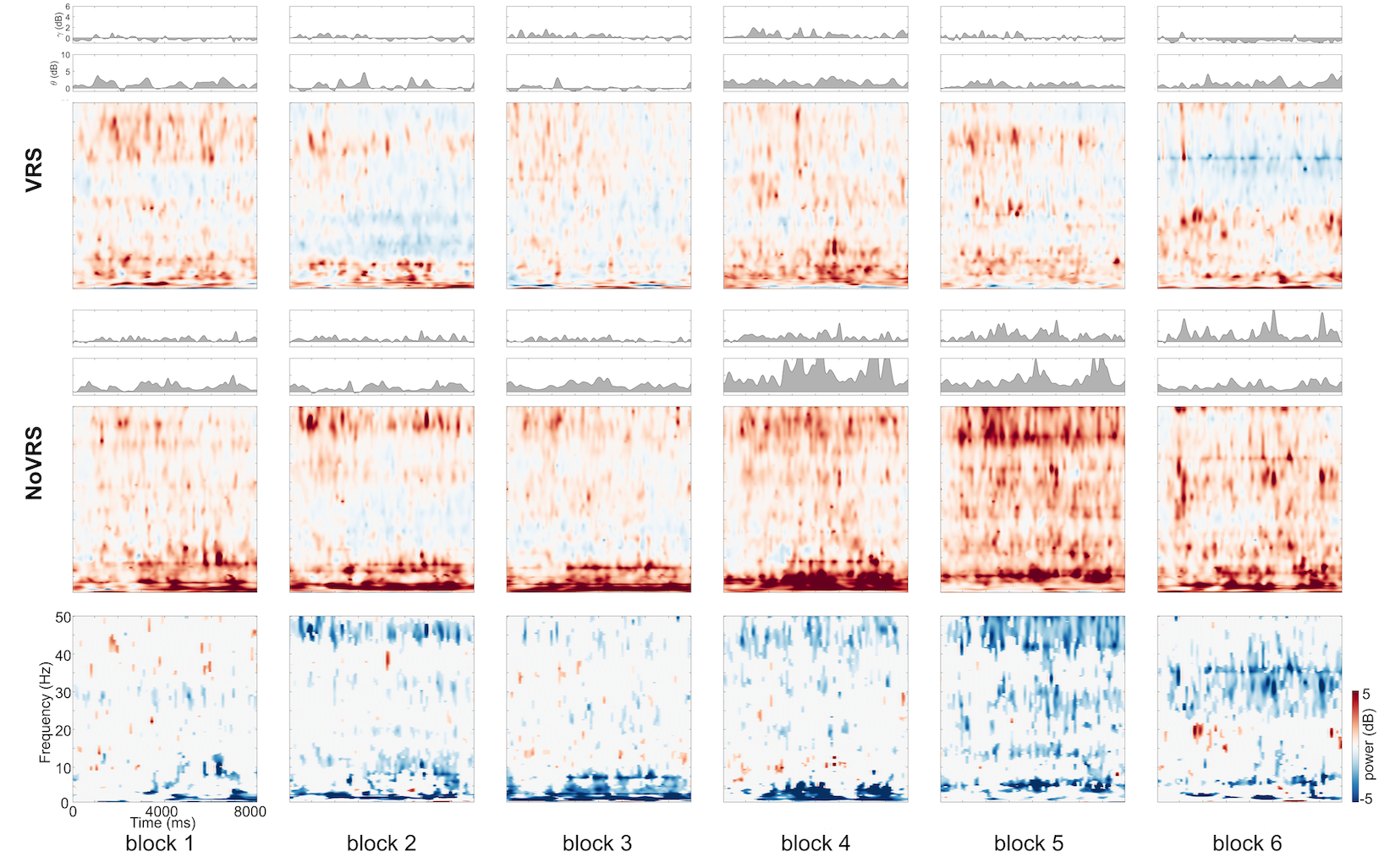}
    \caption{The Event-Related Spectral Dynamics (ERSP) of the \textbf{Cz} channel. The top subplot shows the ERSP of the \textit{VRS} group, the middle shows the ERSP of the \textit{NoVRS} group, and the bottom shows the statistical difference ($p < 0.05$) between the \textit{VRS} and \textit{NoVRS} groups.}
    \label{fig:cz_ersp} 
\end{figure*}

\begin{figure*}[h]
    \centering
    \includegraphics[width=.8\textwidth]{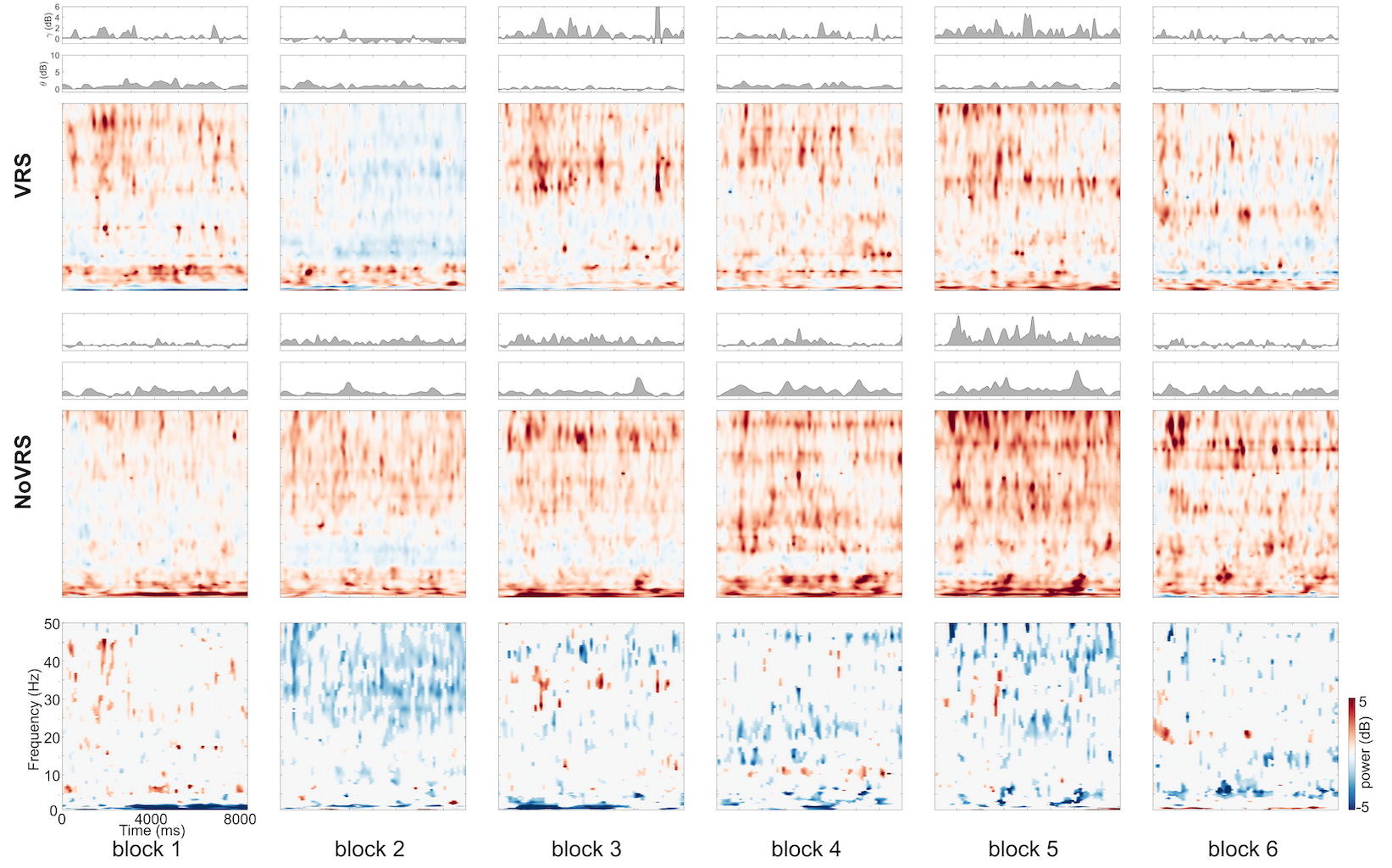}
    \caption{The Event-Related Spectral Dynamics (ERSP) of the \textbf{Pz} channel. The top subplot shows the ERSP of the \textit{VRS} group, the middle shows the ERSP of the \textit{NoVRS} group, and the bottom shows the statistical difference ($p < 0.05$) between the \textit{VRS} and \textit{NoVRS} groups.}
    \label{fig:pz_ersp} 
\end{figure*}

\begin{figure*}[h]
    \centering
    \includegraphics[width=.8\textwidth]{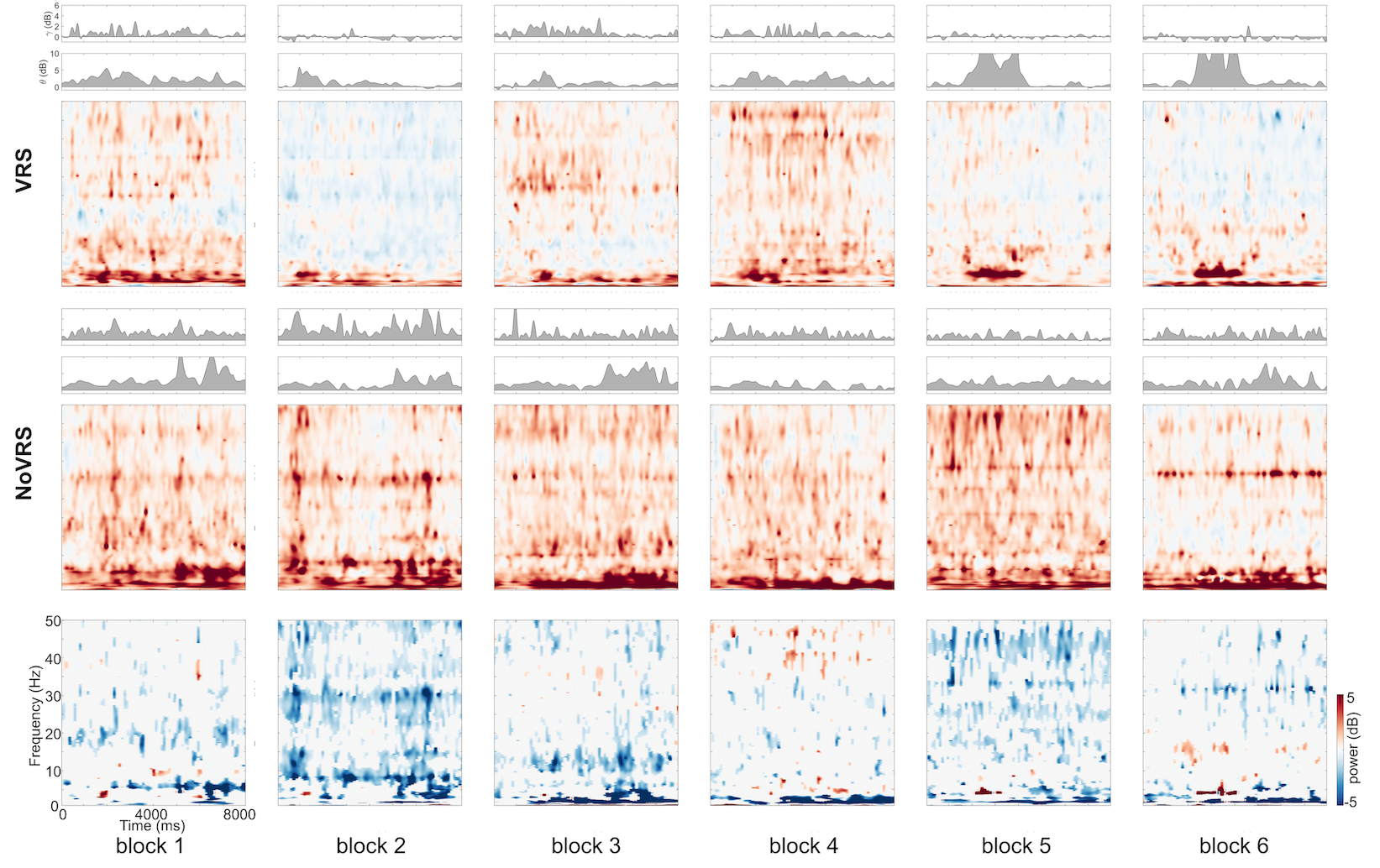}
    \caption{The Event-Related Spectral Dynamics (ERSP) of the \textbf{T8} channel. The top subplot shows the ERSP of the \textit{VRS} group, the middle shows the ERSP of the \textit{NoVRS} group, and the bottom shows the statistical difference ($p < 0.05$) between the \textit{VRS} and \textit{NoVRS} groups.}
    \label{fig:t8_ersp}  
\end{figure*}

%% FOR REPRESENTATIVE PARTICIPANT

%% VRS
\begin{figure*}[h]
    \centering
    \includegraphics[width=\textwidth]{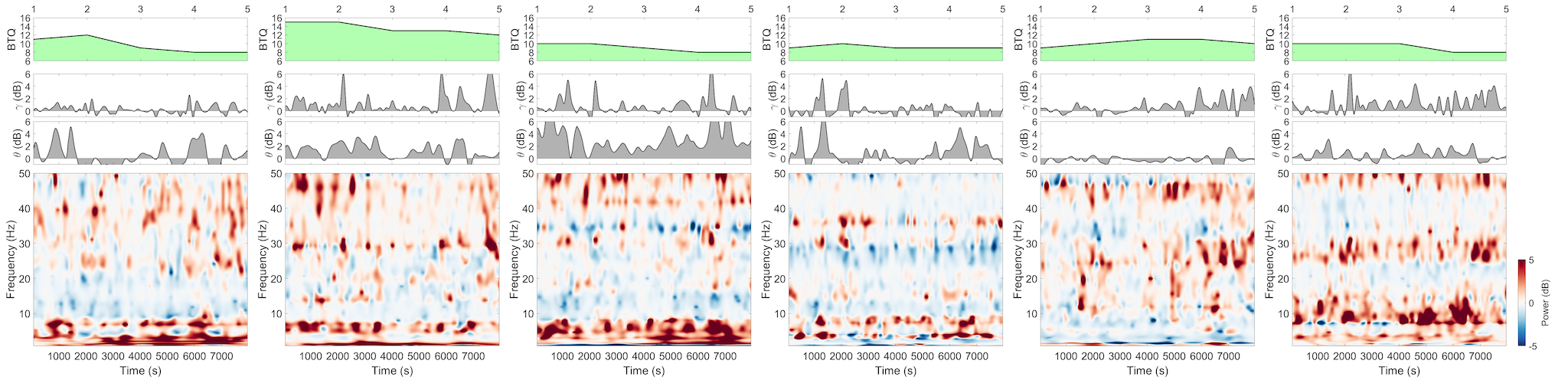}
    \caption{A representative \textit{VRS} participant. The Event-Related Spectral Dynamics (ERSP) of the \textbf{Fz} channel from block 1 (first column) to block 6 (last column). The top subplot shows the value of the Between-Trial Questionnaire (BTQ), the middle shows the EEG power (dB) in the $\gamma$ and $\theta$ band, and the bottom shows the ERSP in the time-frequency domain.}
    \label{fig:Fz_ersp_s07}  
\end{figure*}

\begin{figure*}[h]
    \centering
    \includegraphics[width=\textwidth]{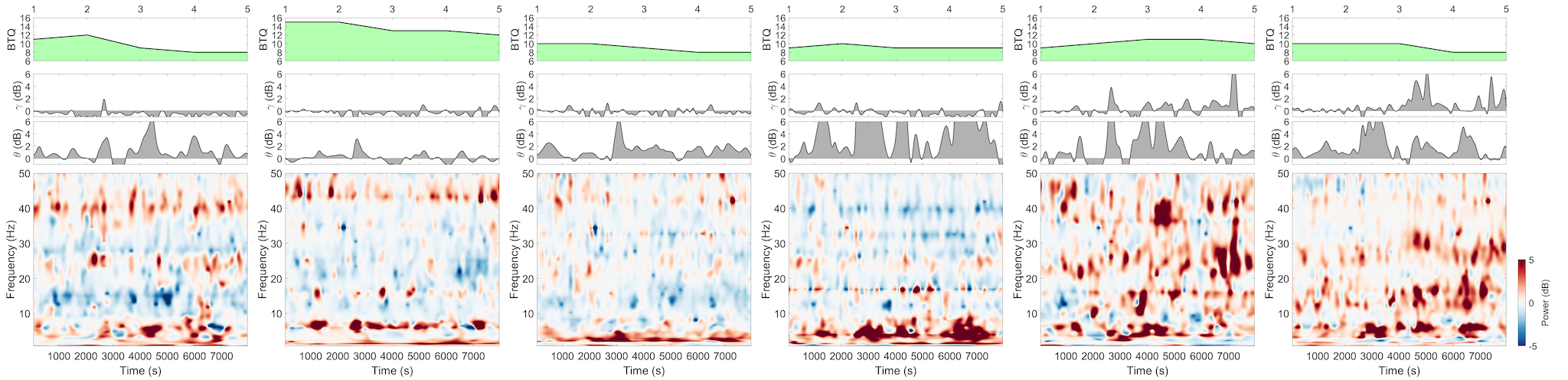}
    \caption{A representative \textit{VRS} participant. The Event-Related Spectral Dynamics (ERSP) of the \textbf{Cz} channel from block 1 (first column) to block 6 (last column). The top subplot shows the value of the Between-Trial Questionnaire (BTQ), the middle shows the EEG power (dB) in the $\gamma$ and $\theta$ band, and the bottom shows the ERSP in the time-frequency domain.}
    \label{fig:Cz_ersp_s07}  % I reuse the same label as previous label
\end{figure*}

\begin{figure*}[h]
    \centering
    \includegraphics[width=\textwidth]{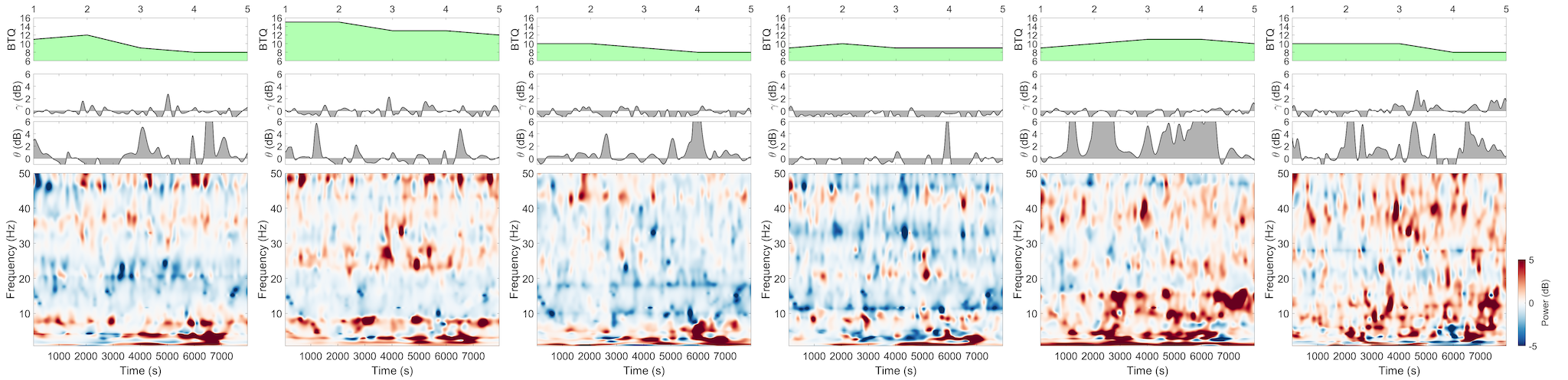}
    \caption{A representative \textit{VRS} participant. The Event-Related Spectral Dynamics (ERSP) of the \textbf{Pz} channel from block 1 (first column) to block 6 (last column). The top subplot shows the value of the Between-Trial Questionnaire (BTQ), the middle shows the EEG power (dB) in the $\gamma$ and $\theta$ band, and the bottom shows the ERSP in the time-frequency domain.}
    \label{fig:Pz_ersp_s07}  % I reuse the same label as previous label
\end{figure*}

\begin{figure*}[h]
    \centering
    \includegraphics[width=\textwidth]{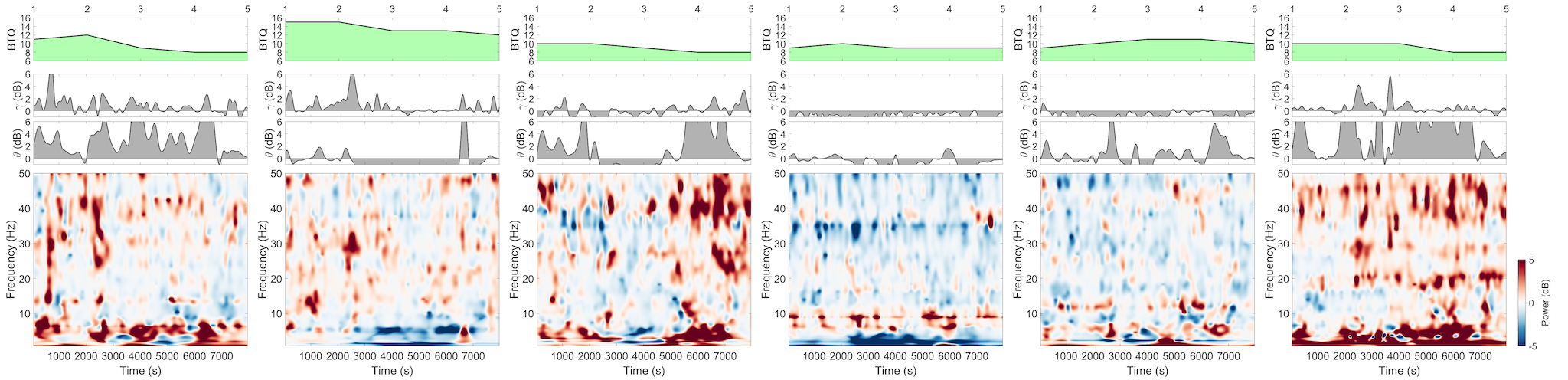}
    \caption{A representative \textit{VRS} participant. The Event-Related Spectral Dynamics (ERSP) of the \textbf{T8} channel from block 1 (first column) to block 6 (last column). The top subplot shows the value of the Between-Trial Questionnaire (BTQ), the middle shows the EEG power (dB) in the $\gamma$ and $\theta$ band, and the bottom shows the ERSP in the time-frequency domain.}
    \label{fig:T8_ersp_s07}  % I reuse the same label as previous label
\end{figure*}

%% NoVRS

\begin{figure*}[h]
    \centering
    \includegraphics[width=\textwidth]{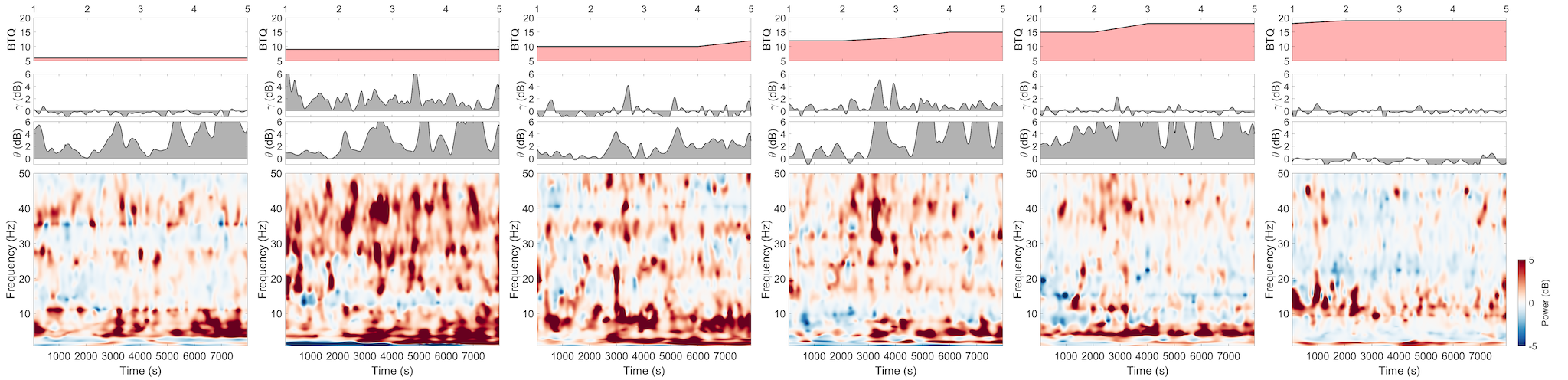}
    \caption{A representative \textit{NoVRS} participant. The Event-Related Spectral Dynamics (ERSP) of the \textbf{Fz} channel from block 1 (first column) to block 6 (last column). The top subplot shows the value of the Between-Trial Questionnaire (BTQ), the middle shows the EEG power (dB) in the $\gamma$ and $\theta$ band, and the bottom shows the ERSP in the time-frequency domain.}
    \label{fig:Fz_ersp_s14}  % I reuse the same label as previous label
\end{figure*}

\begin{figure*}[h]
    \centering
    \includegraphics[width=\textwidth]{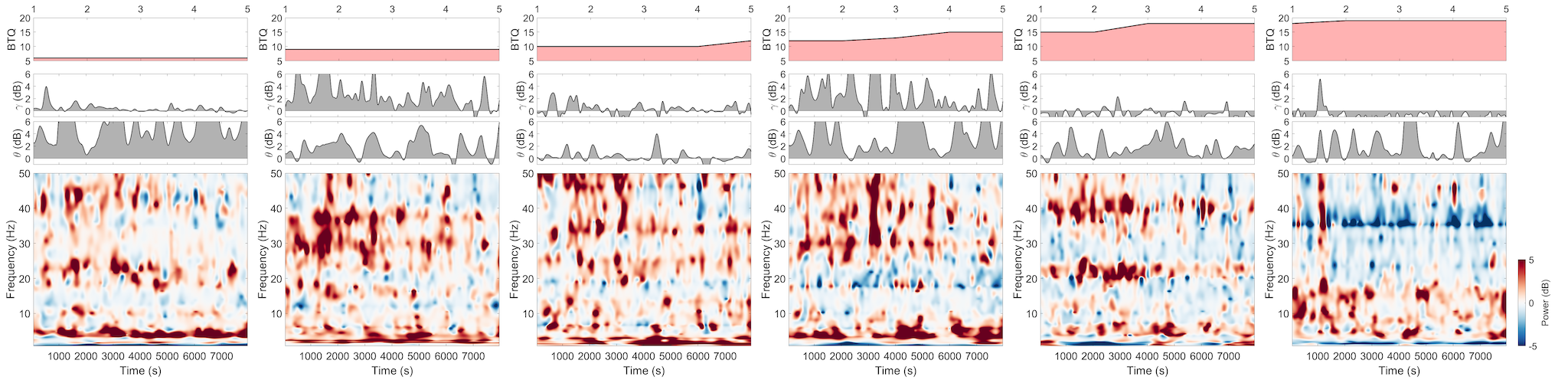}
    \caption{A representative \textit{NoVRS} participant. The Event-Related Spectral Dynamics (ERSP) of the \textbf{Cz} channel from block 1 (first column) to block 6 (last column). The top subplot shows the value of the Between-Trial Questionnaire (BTQ), the middle shows the EEG power (dB) in the $\gamma$ and $\theta$ band, and the bottom shows the ERSP in the time-frequency domain.}
    \label{fig:Cz_ersp_s14}  % I reuse the same label as previous label
\end{figure*}

\begin{figure*}[h]
    \centering
    \includegraphics[width=\textwidth]{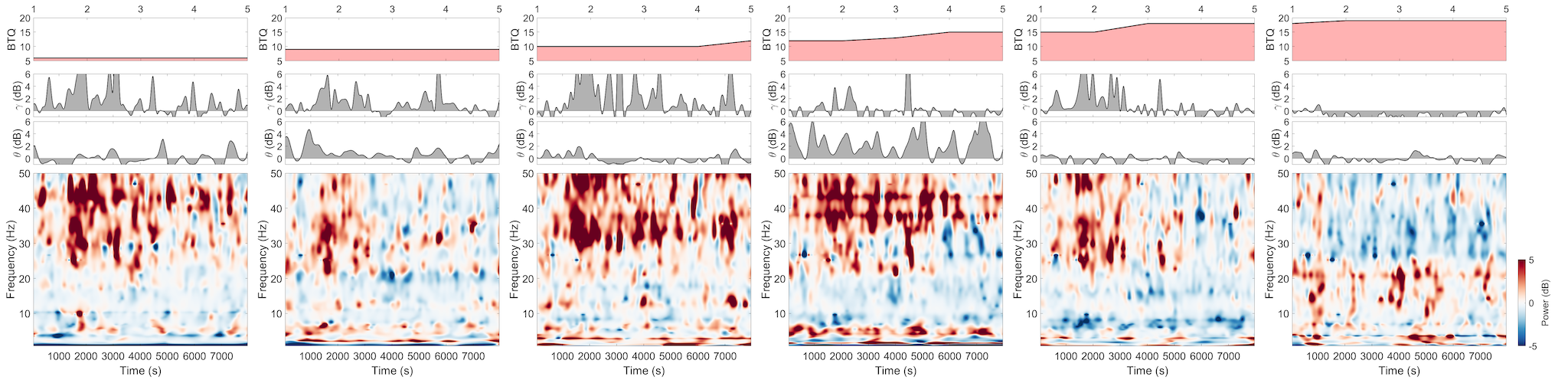}
    \caption{A representative \textit{NoVRS} participant. The Event-Related Spectral Dynamics (ERSP) of the \textbf{Pz} channel from block 1 (first column) to block 6 (last column). The top subplot shows the value of the Between-Trial Questionnaire (BTQ), the middle shows the EEG power (dB) in the $\gamma$ and $\theta$ band, and the bottom shows the ERSP in the time-frequency domain.}
    \label{fig:Pz_ersp_s14}  % I reuse the same label as previous label
\end{figure*}

\begin{figure*}[h]
    \centering
    \includegraphics[width=\textwidth]{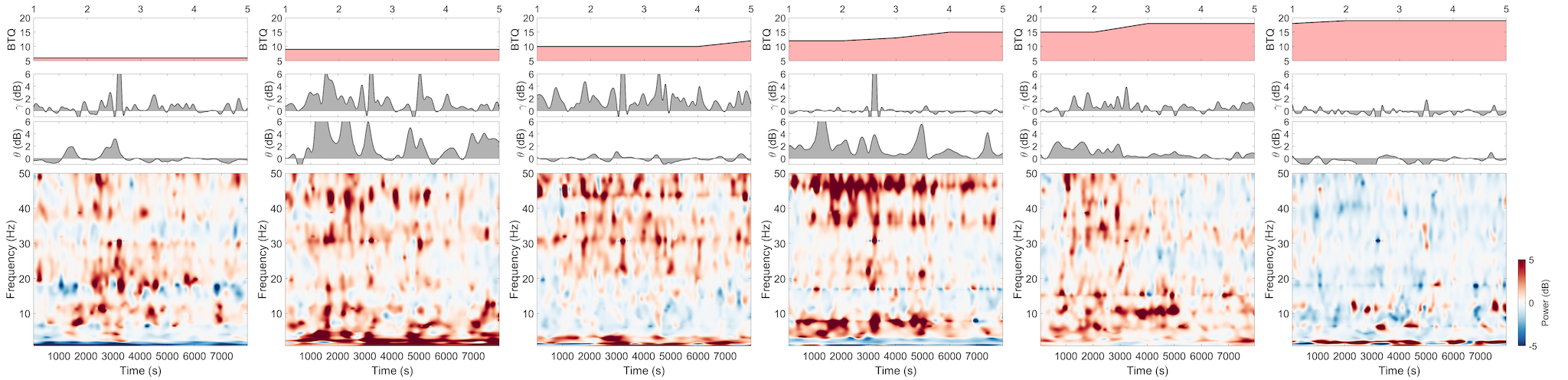}
    \caption{A representative \textit{NoVRS} participant. The Event-Related Spectral Dynamics (ERSP) of the \textbf{T8} channel from block 1 (first column) to block 6 (last column). The top subplot shows the value of the Between-Trial Questionnaire (BTQ), the middle shows the EEG power (dB) in the $\gamma$ and $\theta$ band, and the bottom shows the ERSP in the time-frequency domain.}
    \label{fig:T8_ersp_s14}  % I reuse the same label as previous label
\end{figure*}